\documentclass[prx, twocolumn, superscriptaddress]{revtex4-2}
\usepackage{bm, amsmath, amsfonts, amssymb, braket, mathtools, amsthm}
\usepackage{subfigure}
\usepackage{color}
\usepackage{graphicx}
\usepackage{KSpkg}
\usepackage{dcolumn} % Align table columns on the decimal point
\usepackage{comment}

\usepackage{docmute}

\newtheorem{Definition}{Definition}
\newtheorem{Theorem}{Theorem}
\newtheorem{Proposition}{Proposition}

\newcommand{\whatistitle}{General criterion for non-Hermitian skin effects and Application:\\ Fock space skin effects in many body systems}

\begin{document}

\title{\whatistitle}
\author{Kenji Shimomura}
    \email{kenji.shimomura@yukawa.kyoto-u.ac.jp}
	\affiliation{Center for Gravitational Physics and Quantum Information, Yukawa Institute for Theoretical Physics, Kyoto University, Kyoto 606-8502, Japan}
\author{Masatoshi Sato}
    \email{msato@yukawa.kyoto-u.ac.jp}
	\affiliation{Center for Gravitational Physics and Quantum Information, Yukawa Institute for Theoretical Physics, Kyoto University, Kyoto 606-8502, Japan}
\date{\today}

\begin{abstract}
Non-Hermiticity enables macroscopic accumulation of bulk states, named non-Hermitian skin effects. The non-Hermitian skin effects are well established for single-particle systems, 
but their proper characterization for general systems is elusive. Here, we propose a general criterion of non-Hermitian skin effects, which works for any finite-dimensional system evolved by a linear operator. The applicable systems include many-body systems and network systems.
A system meeting the criterion exhibits enhanced non-normality of the evolution operator, accompanied by exceptional characteristics intrinsic to non-Hermitian systems.
Applying the criterion, we discover a new type of non-Hermitian skin effect in many-body systems, which we dub the Fock space skin effect. We also discuss the Fock space skin effect-induced slow dynamics, which gives an experimental signal for the Fock space skin effect.
\end{abstract}

\maketitle

\paragraph*{Introduction.---}
Localization is a fundamental phenomenon in condensed matter physics.
Disorders or imperfections induce so-called weak localization and Anderson localization,
which form the foundation of quantum transport \cite{Anderson1958,Evers2008}.  
The Pauli exclusion principle and the Coulomb repulsion also result in another localization, Mott localization, enriching quantum phases of strongly correlated systems \cite{Mott2004}.

Recently, a novel class of localization has been extensively studied: 
Non-Hermiticity of the Hamiltonian \cite{Ashida2020,Bergholtz2021,Okuma2023,Bender1998,Bender2002,Bender2007,Rudner2009,Sato2012,Esaki2011, Hu2011,Schomerus2013,Konotop2016,Shen2018,El-Ganainy2018,Gong2018,Kawabata2019,Kawabata2019b,Poli2015,Zeuner2015,Zhen2015,Zhou2018,Weimann2017,Xiao2017,St-Jean2017,Bahari2017,Harari2018,Bandres2018,Zhao2019,Xiao2020} gives rise to a macroscopic accumulation of bulk states on boundaries. 
The new localization phenomenon is called the non-Hermitian skin effect 
\cite{Lee2016,Yao2018,Alvarez2018,Yao2018b,Kunst2018,Yokomizo2019,Kawabata2020,Okuma2019,Edvardsson2019,Lee2019,Borgnia2020,Zhang2022b,Lin2023,Zhu2024,Yoshida2020,Okugawa2020,Kawabata2020b,Fu2021,Song2019,Haga2021,Yang2022,Hwang2023,Zhang2022,Li2020}.
The origin of the non-Hermitian skin effect in terms of topology is well established for single-particle systems: The spectral winding number or the one-dimensional $\mathbb{Z}_2$ number under the periodic boundary condition is responsible for the non-Hermitian skin effects \cite{Okuma2020,Zhang2020}. 
Furthermore, experimental confirmation of the single-particle non-Hermitian skin effect has been done in various systems \cite{Brandenbourger2019,Ghatak2020,Xiao2020,Weidemann2020,Borgnia2020,Hofmann2020,Liang2022,Zhao2023,Ochkan2024,Liu2024}. 

Whereas the non-Hermitian skin effects are also explored in systems other than single-particle ones \cite{Kawabata2022,Alsallom2022,Lee2021,Zheng2024,Mu2020,Lee2020,Cao2023,Feng2024,Haga2021,Kim2024,Li2023,Longhi2023,Hamanaka2023,Faugno2022,Liu2023,Suthar2022,Dora2022,Zhang2022c,Hamazaki2019,Zhang2020b,Wang2022,Mao2023,Wang2022b,Yoshida2024,Gliozzi2024,Hamanaka2024,Kim2023,Wang2023b,Shen2023}, the characterization is less obvious.
No topological description for general cases has been discovered \cite{Kawabata2022};
thus, no clear criterion has been known yet.
The spectrum change between the periodic and open boundary conditions could help detect non-Hermitian skin effects. However, it only captures a partial aspect of them. 
Moreover, precise localization has not been observed occasionally.  
For instance, 
the Pauli exclusion principle prohibits a macroscopic accumulation of bulk states for fermionic many-body systems, obscuring the corresponding non-Hermitian skin effect \cite{Lee2020,Wang2022b,Alsallom2022}.

The purpose of this Letter is to present a general criterion for non-Hermitian skin effects,
which applies to any finite-dimensional system evolved by a linear operator. 
Whereas our criterion is mathematically rigorous, it is based on physical institution.
Systems obeying the criterion reproduce known characteristic features of non-Hermitian skin effects, 
and thus, we can predict related exceptional phenomena based on the criterion.
In particular, by applying the general criterion to many-body systems,  we discover a new type of non-Hermitian skin effect, the Fock space skin effect.

\paragraph*{General criterion for non-Hermitian skin effects.---}

\begin{figure}[htbp]
    \begin{tabular}{cc}
        \begin{minipage}{0.6\hsize}
            \centering
            \includegraphics[width=\columnwidth,clip]{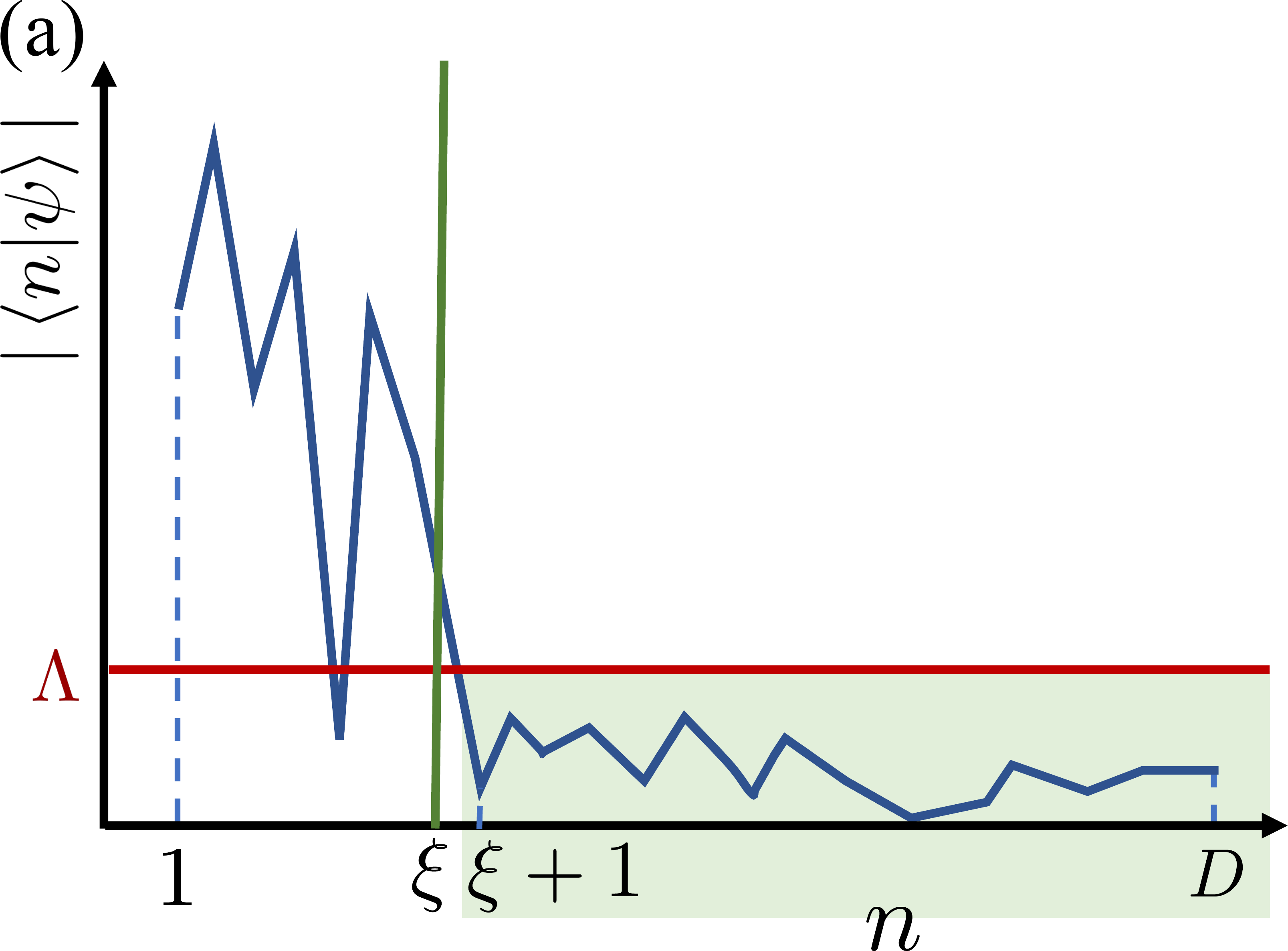}
        \end{minipage}
        &
        \begin{minipage}{0.4\hsize}
            \centering
            \includegraphics[width=\columnwidth,clip]{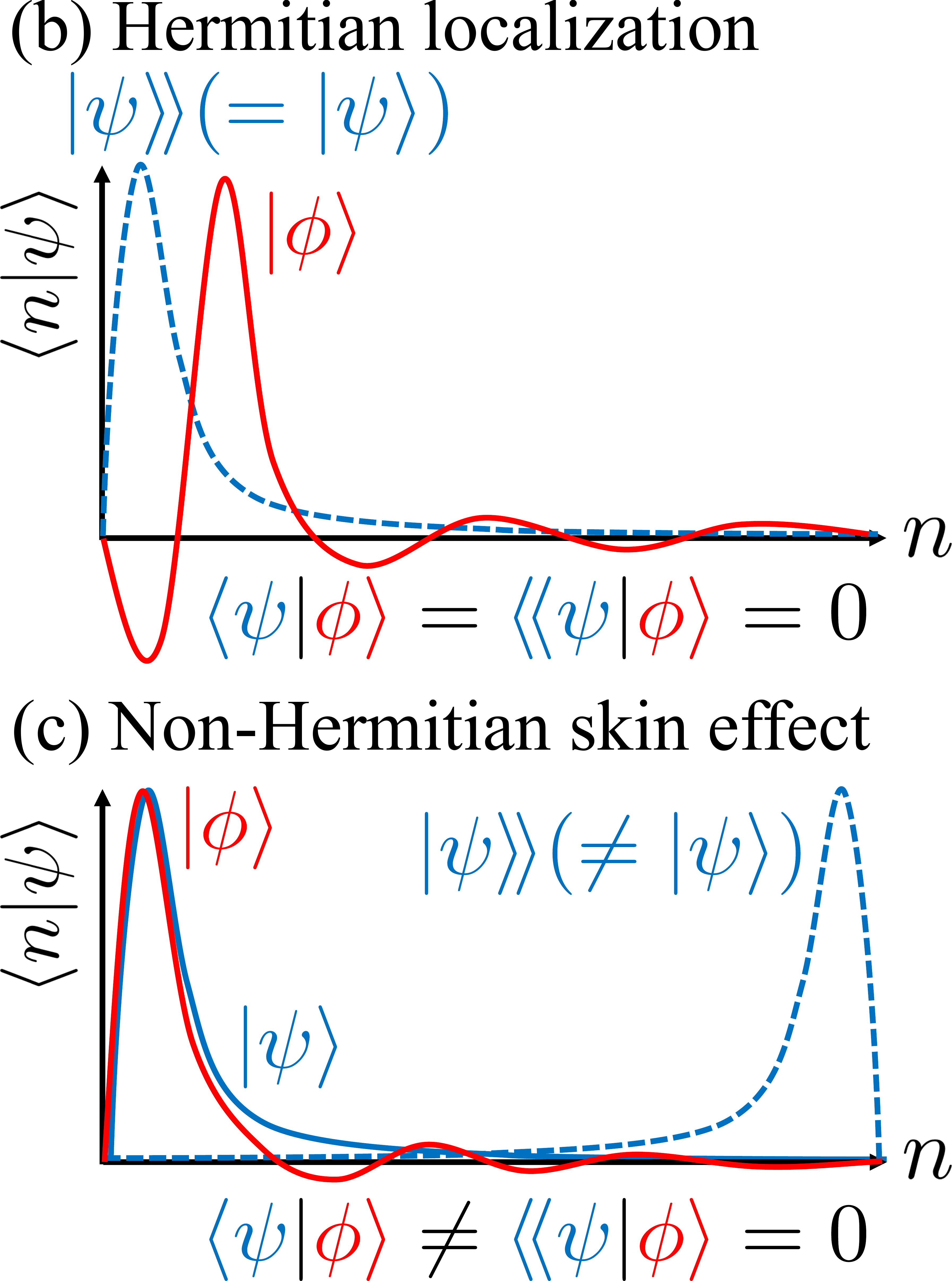}
        \end{minipage}
    \end{tabular}
    \caption{(a) A schematic illustration of the $\Lambda$-localization. The horizontal axis is the label $n$ of the basis state $\ket{n}$, and the vertical one is the amplitude $\abs{\braket{n|\psi}}$ of the state $\ket{\psi}$. 
    The vertical green line represents the localization length $\xi$, and the horizontal red line is the threshold $\Lambda$. 
    (b) Hermitian localization and (c) non-Hermitian skin effect.}
    \label{fig:Lambda-loc}
\end{figure}

Let us start with the definition of localization, which applies to any state in a finite-dimensional Hilbert space,  including a many-body one.
%Below, we use the 2-norm $\norm{\ket{\psi}}\coloneqq \sqrt{\braket{\psi|\psi}}$ as the norm for a state $|\psi\rangle$ and the operator norm $\|\hat{A}\|\coloneqq\min_{\norm{\ket{\phi}}=1}\|\hat{A}\ket{\phi}\|$ as the norm for an operator $\hat{A}$, unless otherwise stated.%For this sake, we consider localization of the expansion coefficient of the vector by a complete orthonormal basis.
Let $\ket{\psi}$ be a state in a Hilbert space $\mathcal{H}$ with the dimension $\dim\mathcal{H}=D>1$, and $\ket{1},\ket{2},\ldots,\ket{D}$ be an orthonormal basis in $\mathcal{H}$.
Then, we define the localization of the state $\ket{\psi}$ as follows:
\begin{Definition}\label{def:1}
For an integer $\xi\in\bkb{1,2,\ldots,D-1}$ and a positive number $\Lambda$, a normalized state $\ket{\psi}$ with $\norm{\ket{\psi}}=\sqrt{\braket{\psi|\psi}}=1$ is defined to be \emph{$\Lambda$-localized} with the localization length $\xi$ when
$
        \abs{\braket{n|\psi}}
        < \Lambda
$
    holds for any $n\in\bkb{\xi+1,\ldots,D}$.
\end{Definition}
We show a schematic illustration of a $\Lambda$-localized state in Fig.~\ref{fig:Lambda-loc} (a).
In the green region where $n$ exceeds the localization length $\xi$, the amplitude $\abs{\braket{n|\psi}}$ is suppressed below the threshold $\Lambda$.
The threshold $\Lambda$ measures the strength of the localization:
A smaller threshold implies more robust localization for a fixed localization length.
This definition is abstract since one can use any orthonormal basis $|n\rangle$.  
If we choose the basis $|i,\sigma\rangle$ labeled by site $i$ with an internal degree of freedom $\sigma$ as the orthonormal basis $|n\rangle$, the above localization becomes conventional. Still, this particular choice is not always necessary for the rigorous characterization of localization inherent in non-Hermitian systems, as discussed below.

Having precisely defined the localization, we now count how many states can accumulate.
Since the localized state is mainly described by $\xi$ basis states $|1\rangle,\dots,|\xi\rangle$ inside the localization length $\xi$, 
a natural expectation is that at most $\xi$ localized states can accumulate for sufficiently small $\Lambda$.
This is true for Hermitian systems. However, due to non-Hermitian skin effects, much stronger accumulation may occur for non-Hermitian systems.

The counterintuitive macroscopic accumulation becomes possible because of the distinction between right and left eigenstates in non-Hermitian systems.
For comparison, let us first consider the Hermitian case. See Fig.\ref{fig:Lambda-loc}(b).
In this case, the right eigenstate $|\psi\rangle$ and the left one $|\psi\rangle\!\rangle$ are the same, and thus, each state is described by a single state, of which the dominant part ensures its independence. 
Therefore, different localized states $|\psi\rangle$ and $|\phi\rangle$ must avoid each other in the restricted Hilbert space within the localization length $\xi$ to keep their independence. 
This exclusion principle limits the possible accumulation number of states by $\xi$.
In contrast, for non-Hermitian systems, we may avoid such a constraint: 
For non-Hermitian systems, the right and left eigenstates are generally different; thus, they can localize differently. In particular, the right localized state $|\psi\rangle$ can overlap with the corresponding left localized state $|\psi\rangle\!\rangle$, mainly in the region outside the localization length, as illustrated in Fig.\ref{fig:Lambda-loc} (c).
(For non-Hermitian skin effects in single-particle systems, the right, and the corresponding left skin modes are indeed localized at opposite boundaries \cite{Okuma2023}.) 
%Hence, they overlap mostly outside the localization length.
Therefore, even when an eigenstate $|\phi\rangle$ is almost identical to $|\psi\rangle$ inside the localization length, it can be orthogonal to the left eigenstate $|\psi\rangle\!\rangle$ simultaneously.
Remarkably, the biorthogonality $\langle\!\langle \psi|\phi\rangle=0$ ensures
the independence of these eigenstates $|\psi\rangle$ and $|\phi\rangle$. 
%the overlap region outside the localization length ensures their independence via the biorthogonal normalization condition between the right and left eigenstates.
Hence, more than $\xi$ right localized states can accumulate together without the contradiction to their independence.
%using the Hilbert space outside the localization length to keep their independence.
The upper limit of the accumulation number of states can be $\mathcal{O}(D)$, which is macroscopically significant in the thermodynamic limit.

The above argument suggests that the presence of $\xi+1$ accumulated localized states gives the minimal condition for non-Hermitian skin effects. More rigorously, we have the following theorem \cite{SM}:
%The concept of the $\Lambda$-localization reveals the disparity of a localization property between Hermitian and non-Hermitian systems.
%The following theorem presents us a fundamental relation between the non-Hermiticity of an operator and the $\Lambda$-localization of eigenvectors of the operator.
%
\begin{Theorem}\label{thm:1}
%    Suppose that the dimension $D$ of the Hilbert space $\mathcal{H}$ is finite.
%    The following holds for any integer $\xi\in\bkb{1,2,\ldots,D-1}$: 
    Let $\hat{H}$ be an operator on $\mathcal{H}$ (such as an effective Hamiltonian, a Liouvillian superoperator, and so on), and $\ket{\psi_1},\ket{\psi_2},\ldots,\ket{\psi_{\xi+1}}\in\mathcal{H}$ be normalized (right) eigenstates of $\hat{H}$ with different eigenvalues.
    Then, if all the $\xi+1$ eigenstates $\ket{\psi_1},\ldots,\ket{\psi_{\xi+1}}$ are $\Lambda$-localized with the threshold $\Lambda\le\dfrac{1}{\sqrt{(\xi+1)(D-\xi)}}\eqqcolon\Lambda_\xi$ for given localization length $\xi$ and orthonormal basis $(|n\rangle)_{n=1,\dots,D}$, then $\hat{H}$ must be non-Hermitian. 
\end{Theorem}

This theorem gives a general criterion for non-Hermitian skin effects: {\it If the system meets the condition of Theorem 1, it shows the non-Hermitian skin effect, namely, localization intrinsic to non-Hermitian systems.}
Note that $\Lambda$-localization depends on the choice and arrangement of the basis $(\ket{n})_n$ as well as the value of $\xi$, but meeting the criterion for a particular basis and $\xi$ is sufficient for the non-Hermitian skin effect.

%One can interpret Theorem~\ref{thm:1} as a No-Go theorem about localization in Hermitian systems.
%By contraposition of Theorem~\ref{thm:1}, $\xi+1$ or more eigenvectors of a Hermitian operator cannnot be $\Lambda$-localized with the localization length $\xi$ simultaneously, when one takes the threshold as $\Lambda=\Lambda_\xi$.
%In short, Hermitian systems involve a trade-off between the localization length $\xi$ and the number of the $\Lambda_\xi$-localized eigenvectors: 
%The number of the $\Lambda_\xi$-localized eigenvectors is never beyond the localization length $\xi$ in Hermitian systems.
%In contrast, non-Hermitian systems can exhibit the localization property beyond such a trade-off:
%The number of the $\Lambda_\xi$-localized eigenvectors can be larger than $\xi$ in non-Hermitian systems.

We have confirmed that the single-particle Hatano-Nelson model \cite{Hatano1996,Hatano1997} satisfies the criterion \cite{SM}. 
Later, we apply the criterion to many body systems and find a new type of skin effect; the Fock space skin effect.

\paragraph*{Non-normality.---}
Non-Hermitian skin effects result in exceptional features:
They give an extreme sensitivity of energy spectra against perturbation, enabling a proposal of a new sensor, {\it i.e.}, the non-Hermitian topological sensor \cite{Budich2020}. 
Furthermore, they also give rise to slowing down the relaxation process without the gap closing \cite{Mori2020,Haga2021}.
Remarkably, these exceptional features in non-Hermitian skin effects originate from non-normality.

An evolution operator $\hat{H}$ is called normal (non-normal) if $\hat{H}$ and its Hermitian conjugation $\hat{H}^\dagger$ commute (do not commute) with each other, {\it i.e.}, $[\hat{H}, \hat{H}^\dagger]= 0$ ($[\hat{H}, \hat{H}^\dagger]\neq 0$).
%The non-Hermitian skin effect in single-particle cases can be understood as enhancement of non-normality of the Hamiltonian 
%, as reviewed below. 
Since a Hermitian operator is always normal, non-normality measures the strength of non-Hermiticity.
A normal operator $\hat{H}$ on ${\cal H}$ is diagonalizable by a unitary matrix, which is equivalent to the condition that the right and left eigenstates coincide.
Therefore, if right and left eigenstates differ, the corresponding evolution operator must be non-normal.
As discussed above, the distinction between right and left eigenstates is essential for the anomalous accumulation of states in non-Hermitian skin effects. Thus, a system showing a non-Hermitian skin effect must be non-normal.
In the proof of Theorem 1, we have shown that a system meeting the criterion in Theorem 1 is not merely non-Hermitian but also non-normal.

To quantify the non-normality of $\hat{H}$, we introduce a useful scalar measure:
%In the presence of the non-Hermitian skin effect, the right and left eigenvectors of the Hamiltonian localize at the opposite edge of each other, which is nothing but a departure from the normality of the Hamiltonian. 
For a diagonalizable $\hat{H}$, we consider the condition number $\kappa(\hat{V})$ \cite{Trefethen2005,Okuma2020b,Nakai2024}
\begin{align}
    \kappa(\hat{V})
    \coloneqq\|\hat{V}\|\|\hat{V}^{-1}\|
    =\sigma_{\rm max}({\hat{V}})/\sigma_{\rm min}(\hat{V})
    \ge 1,
\end{align}
where $\hat{V}$ is an operator diagonalizing $\hat{H}$, $\|\cdot\|$ is the operator two-norm, and $\sigma_{\rm max}(\hat{V})$ and $\sigma_{\rm min}(\hat{V})$ are the maximal and the minimal singular values of $\hat{V}$.
If $\hat{H}$ is normal, $\hat{V}$ is unitary up to an overall constant; thus, the maximal and the minimal singular values are the same, so we have $\kappa(\hat{V})=1$.
On the other hand, if $\hat{H}$ is non-normal, then we have an enhanced condition number $\kappa(\hat{V})>1$.
% When $\hat{A}$ is normal, one can take a unitary operator as $\hat{V}$, resulting in $\kappa(\hat{V})=1$. 
% When $\hat{A}$ is non-normal, conversely, we have $\kappa(\hat{V})\neq 1$.
Therefore, $\kappa(\hat{V})$ measures the non-normality of $\hat{H}$.
For a single-particle system with a non-Hermitian skin effect, one can show that $\kappa(\hat{V})$ grows exponentially with the system length $L$, 
\begin{align}
    \kappa(\hat{V})
    \sim e^{cL} \quad (c>0),
\label{eq:single-particle_kappa}
\end{align}
using the topological origin of the non-Hermitian skin effect \cite{Nakai2024}.
%The Fock space skin effect also enhances the non-normality of an operator as well as the ordinary non-Hermitian skin effect.

In our general setup for non-Hermitian skin effects, no topological description has been known yet, 
so the same topological argument does not work.
Instead, we have the following theorem to estimate the lower bound of $\kappa(\hat{V})$ \cite{SM}.
    
\begin{Theorem}\label{thm:2}
    Let $\hat{H}$ be a diagonalizable operator on $\mathcal{H}$ and $\ket{\psi_1},\ket{\psi_2},\ldots,\ket{\psi_{\xi+1}}$ be
    normalized (right) eigenstates of $\hat{H}$.
    Then, if all the $\xi+1$ eigenstates $\ket{\psi_1},\ket{\psi_2},\ldots,\ket{\psi_{\xi+1}}$ 
    are $\Lambda$-localized with the threshold $\Lambda\le\dfrac{1}{\sqrt{D-\xi}}$,
    the condition number $\kappa(\hat{V})$ of the operator $\hat{V}$ diagonalizing $\hat{H}$ has the lower bound
    \begin{align}
        \kappa(\hat{V})
        > \sqrt{\frac{1}{(\xi+1)(D-\xi)\Lambda^2}-\frac{1}{\xi+1}}
        = \sqrt{\frac{\Lambda_\xi^2}{\Lambda^2}-\frac{1}{\xi+1}}
    \end{align}
\end{Theorem}

It should be noted that the lower bound in Theorem 2 is not optimal:
The highest threshold $\Lambda_\xi$ in Theorem 1 gives a meaningless bound less than 1.
Nevertheless, Theorem~\ref{thm:2} is practically useful because most non-Hermitian skin effects have much lower values of $\Lambda$.
Moreover, without detailed information on eigenstates of $\hat{H}$, one can quickly evaluate the bound since only two parameters, $\xi$ and $\Lambda$, determine the bound. 

The enhanced condition number has various physical implications. It is directly related to the Petermann factor \cite{Petermann1979,Siegman1989}, which quantifies the linewidth broadening resulting from quantum excess noise in lasers and laser gyroscopes. Furthermore, it fairly sharpens the spectral sensitivity against perturbations due to the Bauer-Fike theorem \cite{Trefethen2005, Nakai2024}. It also makes the relaxation time $\tau_0$ longer by $\tau_0\log \kappa(\hat{V})$ in the transit dynamics, as described below.

\paragraph*{Fock space skin effect.---}

A significant advantage of our general criterion is flexibility in choosing the orthonormal basis.
The flexibility enables us to characterize non-Hermitian skin effects in many-body systems properly. 
For many-body systems, the basis in the real space coordinate has multiple site indices $|i_1,i_2 \dots \rangle$, where $i_p$ is the site of $p$th particle. Hence, the characterization of the non-Hermitian skin effect is not apparent, in contrast to the single-particle case.
The average position of particles could help examine the many-body non-Hermitian skin effect, but this characterization severely restricts information obtained for many-body states. 
Keeping an application to interacting systems in mind, we choose the Fock space as the orthogonal basis. We propose the concept of the Fock space skin effect as a localization phenomenon inherent in non-Hermitian many-body systems.

Let $\hat{H}$ be an operator on a finite-dimensional Fock space $\mathcal{F}$ (such as fermionic, hard-core bosonic, or spin systems on a finite lattice) with the dimension $D>1$, and $(\ket{n})_{n=1}^D$ be orthonormal Fock space basis of $\mathcal{F}$.
Then, the exact definition of the Fock space skin effect is as follows.

\begin{Definition}\label{def:2}
If there is an integer $\xi$ such that at least $\xi+1$ eigenstates of $\hat{H}$ are $\Lambda_\xi$-localized with the localization length $\xi$, we call such localization {\rm the Fock space skin effect}.
\end{Definition}
From Theorem~\ref{thm:1} for the Fock space ${\cal F}$, the Fock space skin effect is a non-Hermitian one.

To demonstrate the Fock space skin effect, we consider the fermionic Hatano-Nelson model with interaction:
\begin{align}\label{eq:iHN}
    \hat{H}_{\rm iHN}
    = \sum_{j=1}^{L-1} \bka{e^\alpha\hat{c}_j^\dag\hat{c}_{j+1} + e^{-\alpha}\hat{c}_{j+1}^\dag\hat{c}_j + U\hat{n}_j\hat{n}_{j+1}},
\end{align}
where $\hat{c}_j$ ($\hat{c}_j^\dagger$) is the fermionic annihilation (creation) operator at site $j$ with the anticommutation relations $\{\hat{c}_i,\hat{c}_j\}=\{\hat{c}_i^\dag,\hat{c}_j^\dag\}=0$, $\{\hat{c}_i,\hat{c}_j^\dag\}=\delta_{i,j}$, and $\hat{n}_j=\hat{c}^\dagger_j\hat{c}_j$ is the particle number at site $j$. 

For $\alpha\neq 0$, this model shows the energy spectrum difference between the periodic and open boundary conditions \cite{SM}: The model shows a complex spectrum under the periodic boundary condition but a real spectrum under the open boundary condition. 
Thus, we can expect the Fock space skin effect.
Below, we consider the open boundary condition.

As the total particle number $\hat{N}=\sum_{j=1}^L\hat{n}_j$ commutes with $\hat{H}_{\rm iHN}$, we can restrict the Hilbert space into the $N$-particle fermionic Fock space $\mathcal{F}_N$ of the dimension $\dim\mathcal{F}_N = L!/[N!(L-N)!]$. 
To show the Fock space skin effect, we arrange the order of the Fock basis $(\ket{n})_{n=1}^{\dim\mathcal{F}_N} = (\hat{c}_{j_1}^\dag\cdots\hat{c}_{j_N}^\dag\ket{0}\mid 1\le j_1<\cdots<j_N\le L)$ of $\mathcal{F}_N$ as follows:
We take a reference eigenstate $\ket{\Psi_1}$ of $\hat{H}_{\rm iHN}$, and rearrange the order of the Fock basis $\ket{n}$ to satisfy the inequality,
%\begin{align}
$
    \abs{\braket{1|\Psi_{1}}}
    \ge \abs{\braket{2|\Psi_{1}}}
    \ge \cdots
    \ge \abs{\braket{\dim\mathcal{F}_N|\Psi_{1}}}.
$
%\end{align}
Then, we search a localization length $\xi$ and $\xi+1$ eigenstates $(\ket{\Psi_m})_{m=1,\dots,\xi+1}$ of $\hat{H}_{\rm iNH}$ that satisfy the Definition~\ref{def:2}.

\begin{figure}
    \centering
    \includegraphics[width=\columnwidth,clip]{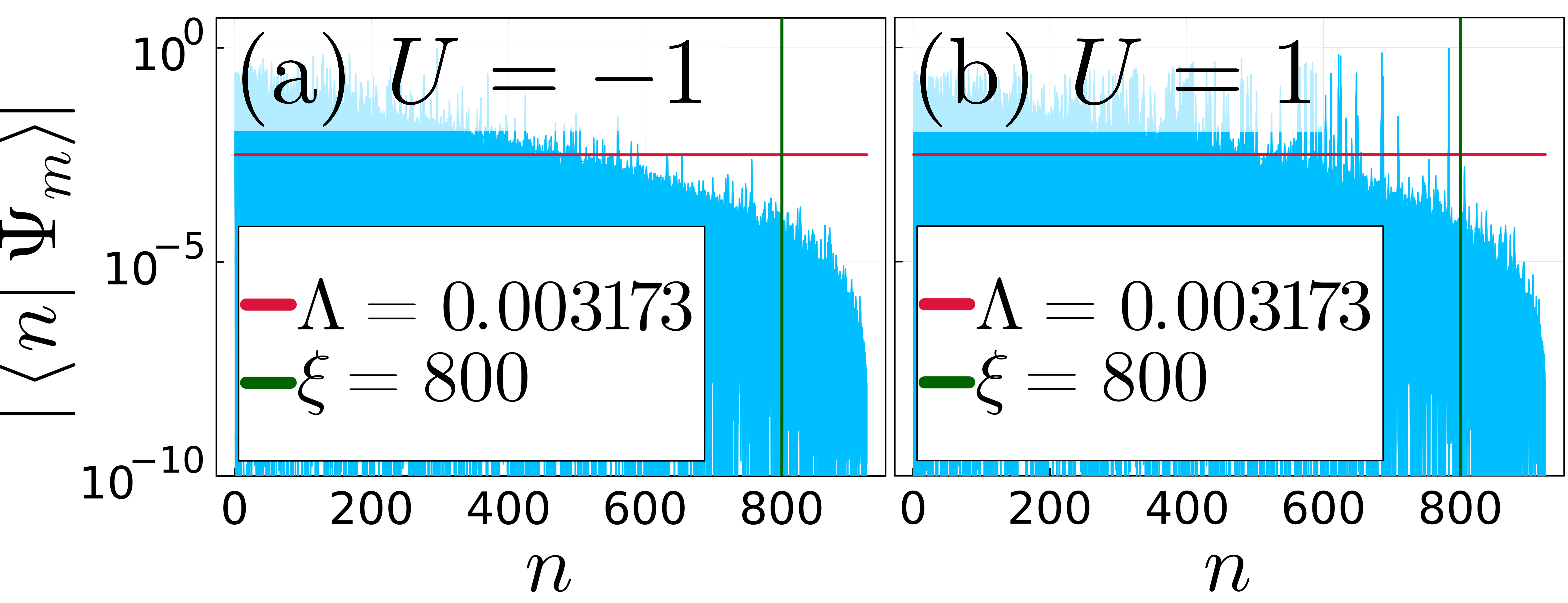}
    \caption{The Fock space skin effect in the fermionic interacting Hatano-Nelson model [$\alpha=0.5$, $L=12$, $N=6$; (a) $U=-1.0$; (b) $U=1.0$]. We take the localization length as $\xi=800$ (the green solid line) and plot just $\xi+1$ eigenstates. 
    One can see the macroscopic accumulation of the eigenstates in the Fock space.
    All of the amplitudes $\abs{\braket{n|\Psi_m}}$ are smaller than the threshold $\Lambda_\xi=0.003\,173$ (the red solid line) for $n\ge\xi+1$, with $D=L!/[N!(L-N)!]=924$, which indicates the Fock space skin effect.}
    \label{fig:interacting_HatanoNelson}
\end{figure}

In Fig.~\ref{fig:interacting_HatanoNelson}, we illustrate the Fock space skin effect obtained in this manner:
$\xi+1$ eigenstates are $\Lambda_\xi$-localized with the localization length $\xi$, so the system meets our criterion.
Both for the attractive $(U<0)$ and repulsive $(U>0)$ interactions, the system shows the Fock space skin effect.

% $\hat{H}$ be an operator on $\mathcal{H}$ such that $\ket{\psi_1},\ldots,\ket{\psi_{\xi+1}}$ are the eigenvectors on $\hat{H}$

As mentioned above, the non-Hermitian skin effect results in the non-normality of the Hamiltonian.
Therefore, the Fock space skin space effect also enhances the non-normality.
Now, using Theorem~\ref{thm:2}, we evaluate the scalar measure of non-normality $\kappa(\hat{V})$ for the interacting Hatano-Nelson model.

\begin{figure}
    \begin{tabular}{cc}
        \begin{minipage}{0.5\hsize}
            \centering
            \includegraphics[width=\columnwidth,clip]{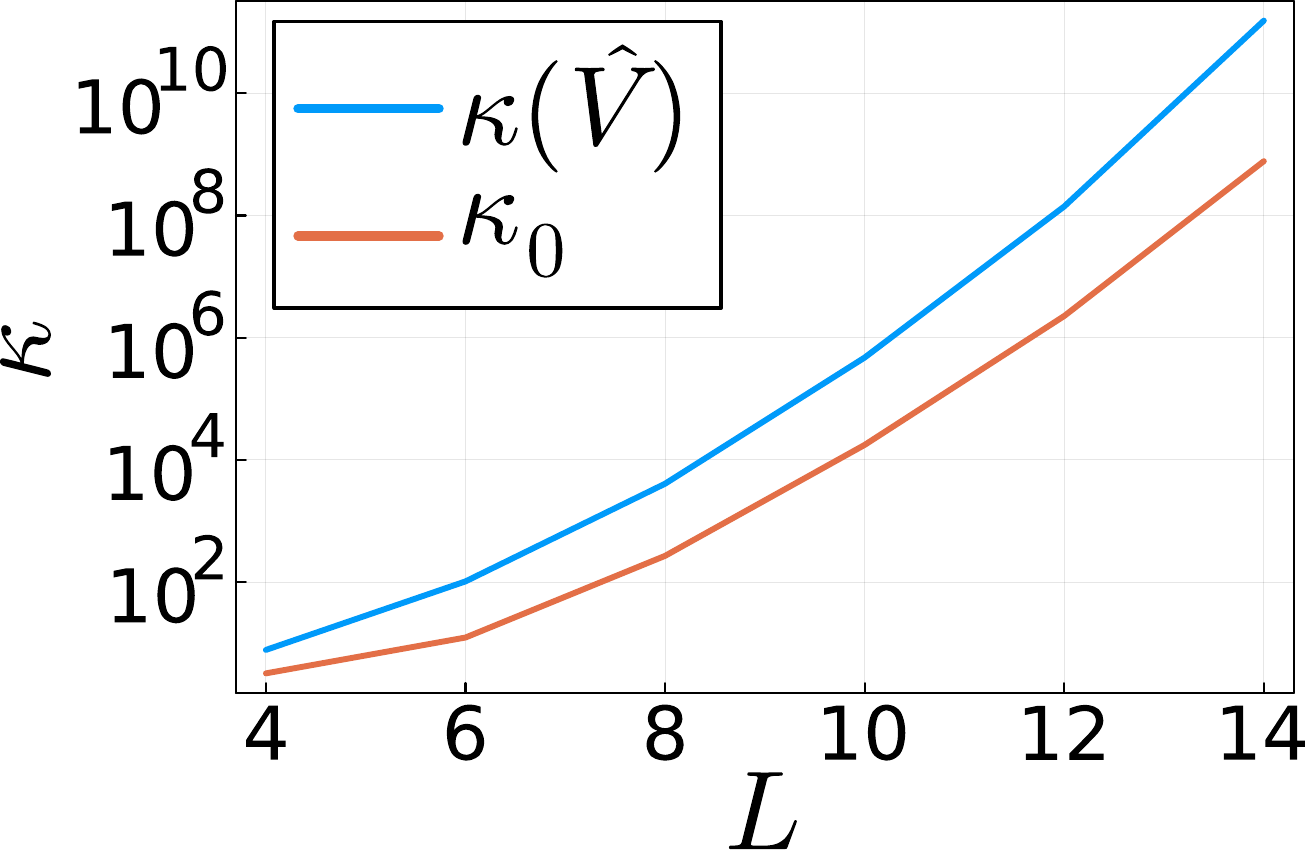}
        \end{minipage}
        &
        \begin{minipage}{0.5\hsize}
            \centering
            \includegraphics[width=\columnwidth,clip]{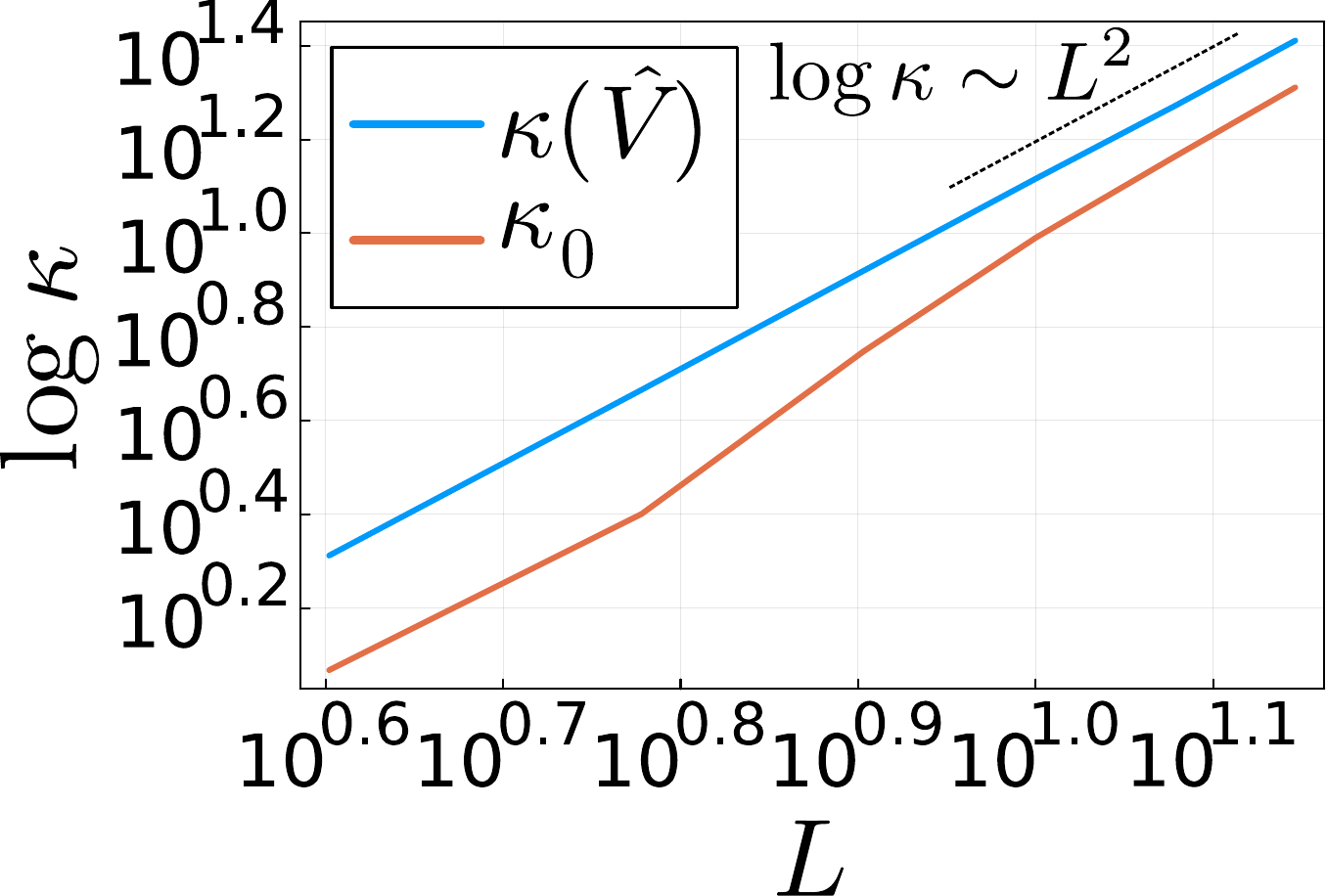}
        \end{minipage}
    \end{tabular}
    \caption{The system size $L$ dependence of the condition number $\kappa(\hat{V})$ and its lower bound $\kappa_0$ in Eq.\eqref{eq:kappa0} for the interacting fermionic Hatano-Nelson model in Eq.\eqref{eq:iHN}. The model parameters are $\alpha=0.5$, $U=-1$, and $N=L/2$.
    The left figure shows that both of $\kappa(\hat{V})$ and $\kappa_0$ increase more rapidly than $e^{\mathcal{O}(L)}$ as $L$ becomes large.
    The right figure indicates that the system size dependence of $\log\kappa(\hat{V})$ and $\log\kappa_0$ approaches $\mathcal{O}(L^2)$.}
    \label{fig:condition_number}
\end{figure}

To obtain a severe bound from Theorem~\ref{thm:2}, we should choose the lowest value of the threshold $\Lambda$. 
We can expect that all states in the interacting Hatano-Nelson model show the Fock space skin effect, so 
we choose $\xi=\dim{\cal F}_N-1$, which gives the lowest value of $\Lambda$.
Because the lowest value of $\Lambda$ obeys
$
\Lambda+\epsilon< {\rm max}_{m=1,\dots,\dim{\cal F}_N}|\braket{\dim {\cal F}_N|\Psi_m}|
$
for $\epsilon= 0_+$, Theorem~\ref{thm:2} leads to the bound
\begin{align}\label{eq:kappa0}
    \kappa(\hat{V})\ge \frac{1}{\sqrt{\dim {\cal F}_N}}\sqrt{\frac{1}{\max_m|\braket{\dim {\cal F}_N|\Psi_m}|^2}-1}   
    \eqqcolon\kappa_0.
\end{align}
Figure \ref{fig:condition_number} compares $\kappa(\hat{V})$ and $\kappa_0$ for the interacting Hatano-Nelson model at the half-filling ($N=L/2$) regarding the fermion $\hat{c}_j^\dag$ in the site basis. 
Remarkably, $\kappa_0$ reproduces the system size dependence of $\kappa(\hat{V})$ for large $L$ correctly, giving a good measure for the enhanced non-normality.
Figure \ref{fig:condition_number} also shows that many-body systems may exhibit much stronger non-normality than single-particle systems:
It shows $\kappa(\hat{V})\sim e^{c'L^2}$ ($c'>0$) for $N=L/2$, which is beyond the linear behavior in $L$ of the exponential in Eq.(\ref{eq:single-particle_kappa}).
Generally, for $L\ge N\ge 1$, we estimate that the interacting Hatano-Nelson model shows \cite{SM} 
\begin{align}
\kappa_0\sim e^{\alpha N(L-N)}/\sqrt{\dim {\cal F}_N}.    
\label{eq:kappa0_eval}
\end{align}
This equation reproduces the $L^2$ dependence in the exponential for $N=L/2$ as well as the $L$ dependence in the exponential in Eq.(\ref{eq:single-particle_kappa}) for $N=1$. It also indicates that whereas the Pauli exclusion principle obscures the particle accumulation, the Fock space skin effect occurs except for $N=L$, where particles cannot move anymore due to the Pauli exclusion principle.
The superexponential non-normality crucially sharpens the spectral sensitivity to perturbations via the Bauer-Fike theorem \cite{Trefethen2005}.

\paragraph*{Slow relaxation due to Fock space skin effect.---}

\begin{figure}[htbp]
    \begin{tabular}{cc}
        \begin{minipage}{0.5\hsize}
            \centering
            \includegraphics[width=\columnwidth,clip]{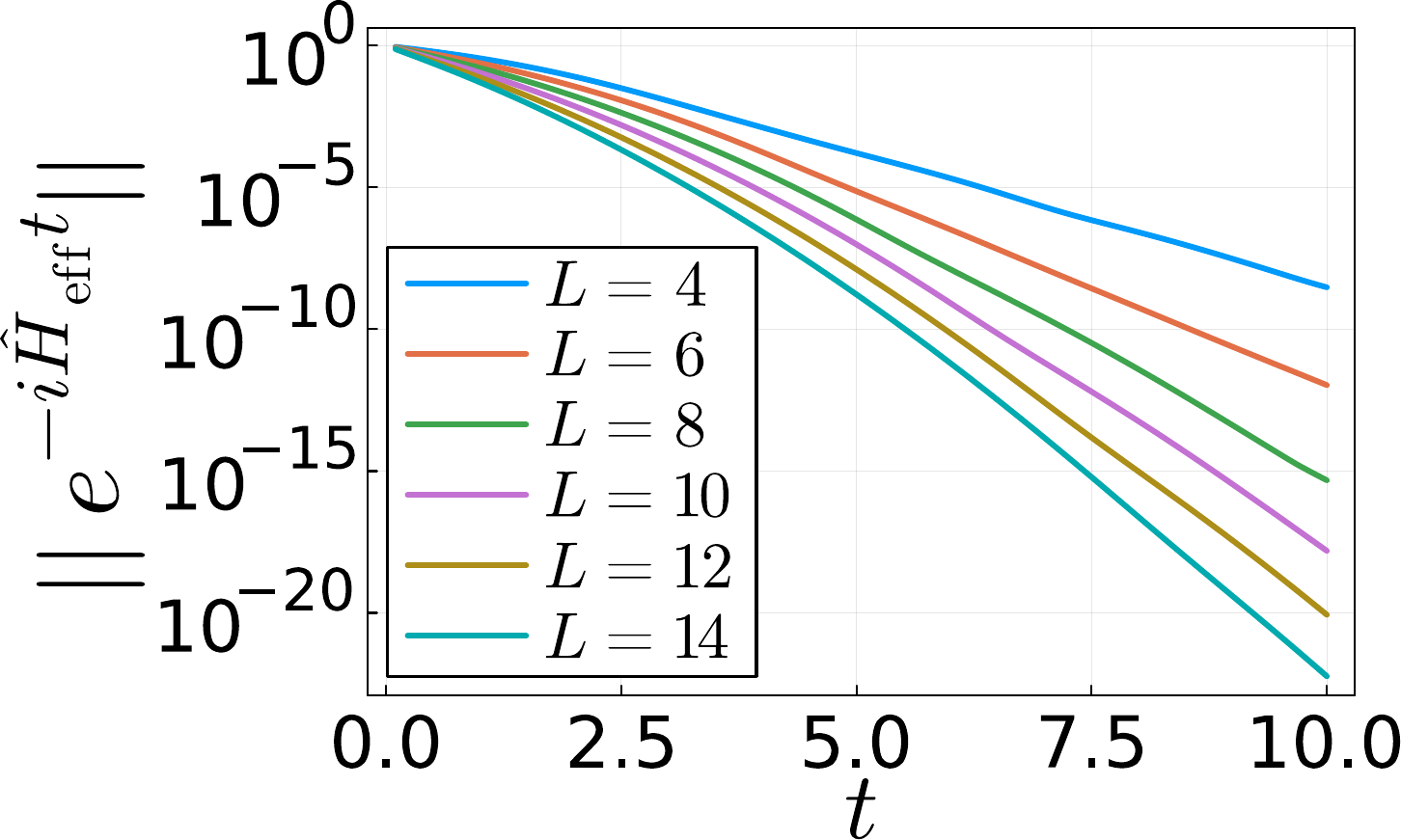}
        \end{minipage}
        &
        \begin{minipage}{0.5\hsize}
            \centering
            \includegraphics[width=\columnwidth,clip]{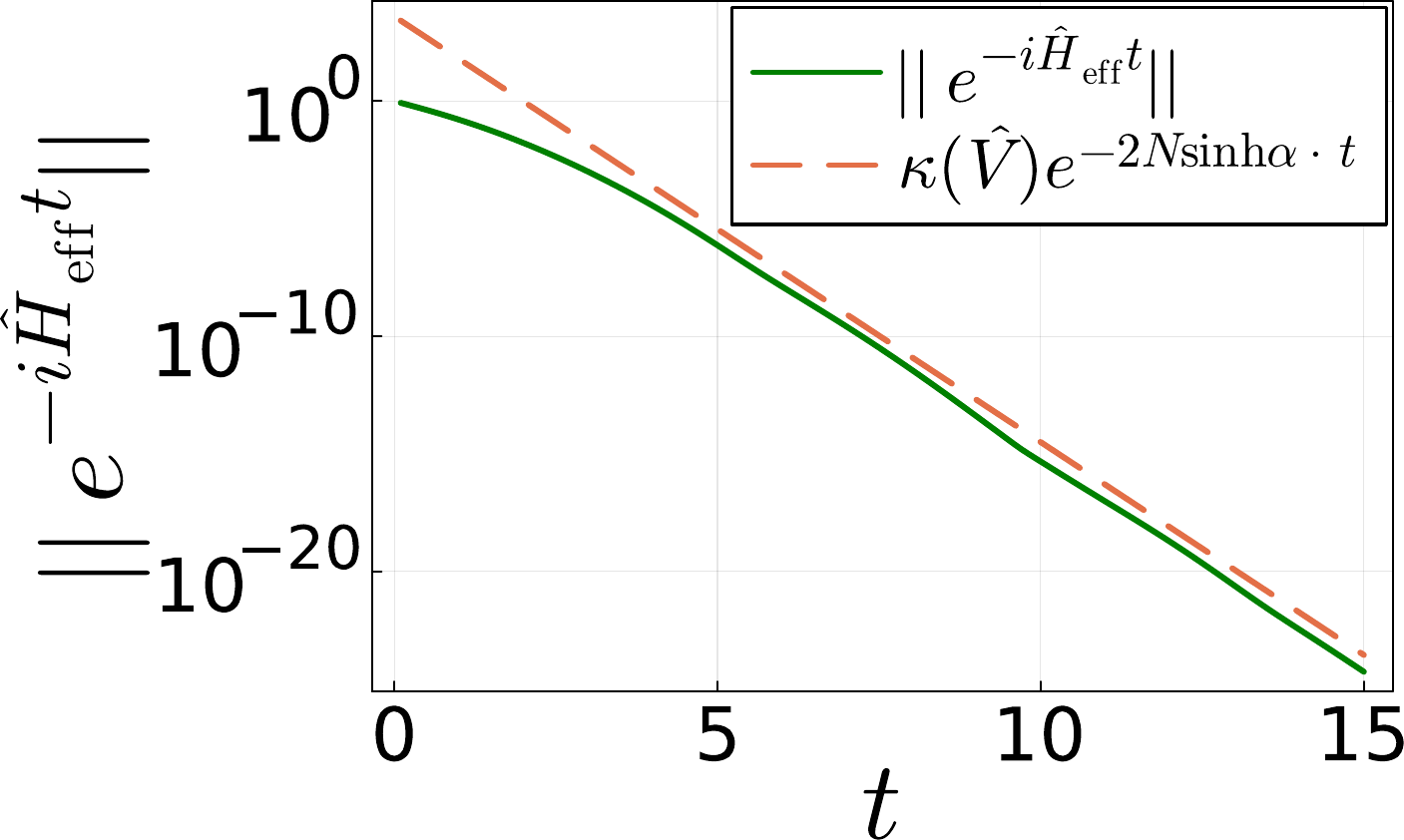}
        \end{minipage}
    \end{tabular}
    \caption{The nonunitary dynamics $\|e^{-i\hat{H}_{\mathrm{eff}}t}\|$ of the effective Hamiltonian $\hat{H}_{\mathrm{eff}}$ in Eq.\eqref{eq:effectiveHam} [$\alpha=0.5$, $U=-1.0$, $N=L/2$ (half-filling); right side: $L=8$, $N=4$]. 
    }
    \label{fig:nonunitary_dynamics}
\end{figure}

The enhanced non-normality due to the Fock space skin effect results in the slowing down of the relaxation time.
As mentioned above, this is a general consequence of the non-Hermitian skin effect. For concreteness, we examine a particular process showing this feature here.
Let us consider an open quantum system governed by the Lindblad equation \cite{Lindblad1976,Gorini1976}
\begin{align}\label{eq:Lindblad}
    \frac{\partial\hat{\rho}}{\partial t}
    &= -i(\hat{H}_0\hat{\rho}-\hat{\rho}\hat{H}_0) + \sum_r\left(\hat{L}_r\hat{\rho}\hat{L}_r^\dag - \frac{1}{2}\{\hat{L}_r^\dag\hat{L}_r,\hat{\rho}\}\right) \nonumber\\
    &= -i(\hat{H}_{\mathrm{eff}}\hat{\rho}-\hat{\rho}\hat{H}_{\mathrm{eff}}^\dag) + \sum_r \hat{L}_r\hat{\rho}\hat{L}_r^\dag, 
\end{align}
where the Hermitian Hamiltonian $\hat{H}_0$ and the Lindblad operator $\hat{L}_r$ are given by
\begin{align}
    &\hat{H}_0
    = \sum_{j=1}^{L-1}\bka{\cosh\alpha (\hat{c}_j^\dag\hat{c}_{j+1} + \hat{c}_{j+1}^\dag\hat{c}_j) + U\hat{n}_j\hat{n}_{j+1}}, \nonumber\\
    &\hat{L}_j
    = \sqrt{2\sinh\alpha}(\hat{c}_j+i\hat{c}_{j+1})
    \quad
    (j=1,2,\ldots,L-1), \nonumber\\
    &\hat{L}_0
    = \sqrt{2\sinh\alpha}\hat{c}_1,
    \quad
    \hat{L}_L
    = \sqrt{2\sinh\alpha}\hat{c}_L.
    \label{eq:H&L}
\end{align}
Then, the effective Hamiltonian $\hat{H}_{\mathrm{eff}}$ realizes the fermionic Hatano-Nelson model with interaction \cite{Gong2018}
\begin{align}\label{eq:effectiveHam}
    \hat{H}_{\mathrm{eff}}
    = \hat{H}_0 - \frac{i}{2}\sum_{r}\hat{L}_r^\dag\hat{L}_r %\nonumber\\
    =\hat{H}_{\rm iHN} 
    %\sum_{j=1}^{L-1} \bka{e^\alpha\hat{c}_j^\dag\hat{c}_{j+1} + e^{-\alpha}\hat{c}_{j+1}^\dag\hat{c}_j 
    %+ U\hat{n}_j\hat{n}_{j+1}}
    - 2i\sinh\alpha\hat{N}.
\end{align}
The slowing down due to the Fock space skin effect becomes evident for 
the survival probability $P(t;\hat{\rho}_{\mathrm{init}})$ that the total particle number does not decrease during the time interval $t$ for the initial state.
The probability $P(t;\hat{\rho}_{\mathrm{init}})$ is of the form \cite{SM},
$
    P(t;\hat{\rho}_{\mathrm{init}})
    = \tr\bka{e^{-i\hat{H}_{\mathrm{eff}}t}\hat{\rho}_{\mathrm{init}}e^{i\hat{H}_{\mathrm{eff}}^\dag t}}.
$
In particular, for a pure initial state $\hat{\rho}_{\rm init}=\ket{\Psi_{\rm init}}\bra{\Psi_{\rm init}}$, 
we obtain $P(t;\hat{\rho}_{\rm init})=\|e^{-i\hat{H}_{\mathrm{eff}}t}\ket{\Psi_{\mathrm{init}}}\|^2$, which is governed by the norm of the evolution operator $\|e^{-i\hat{H}_{\mathrm{eff}}t}\|$.

%As shown in Fig.~\ref{fig:nonunitary_dynamics}, one can see that the decrease in $\|e^{-i\hat{H}_{\mathrm{eff}}t}\|$ for small $t$ is slower than for large $t$.
%The enhanced non-normality, such as slow dynamics for small $t$
%}

Remarkably, the condition number gives the upper bound of $\|e^{-i\hat{H}_{\mathrm{eff}}t}\|$:
For $\hat{V}$ diagonalizing 
$\hat{H}_{\mathrm{eff}}$ (and thus $\hat{H}_{\mathrm{iHN}}$) and
the complex eigenenergies $E_m$ ($m=1,\ldots,\dim\mathcal{F}_N$) of $\hat{H}_{\mathrm{eff}}$,
we have $
    \|e^{-i\hat{H}_{\mathrm{eff}}t}\|
    =\|\hat{V}^{-1}e^{-i{\rm diag}(E_1,E_2,\dots)t}\hat{V}\|
    \le \kappa(\hat{V})e^{\max_m \operatorname{Im}E_m t}
$~\footnote{Note that ${\rm Im}E_m$ is negative since it is given by $-i\sum_r L_r^\dagger L_r/2$ in $H_{\rm eff}$.}.
Then, it holds $\operatorname{Im}E_m = -2N\sinh\alpha$ for any $m$ because $H_{\rm iHN}$ has a real spectrum under the open boundary condition.
As a result, we have
$
    \|e^{-i\hat{H}_{\mathrm{eff}}t}\|\le \kappa(\hat{V})e^{-2Nt\sinh\alpha }.   
$
As shown in Fig. \ref{fig:nonunitary_dynamics}, 
the upper bound $\kappa(\hat{V})e^{-2Nt\sinh\alpha}$ reproduces the asymptotic behavior of $\|e^{-iH_{\rm eff}t}\|$ correctly.

Using the upper bound, we can estimate the relaxation time $t=\tau$ when the survival probability becomes exponentially small as
\begin{align}
\tau= [1+\log \kappa(\hat{V})]\tau_0  
\end{align}
where $\tau_0=(2N\sinh\alpha)^{-1}$ is the conventional relaxation time evaluated from the energy.
Then, from Eq.(\ref{eq:kappa0_eval}), we conclude that the Fock space skin effect exceptionally slows down the relaxation time of the survival probability, which could be observable in experiments \cite{SM}.

\paragraph*{Discussion.---}
In this Letter, we propose a general criterion for non-Hermitian skin effects. Applying it to many-body systems, we discover a novel class of non-Hermitian skin effects, the Fock space skin effect.  
We also discuss a possible experimental signal for the Fock space skin effect.

Whereas we focus on many-body systems in this Letter, our criterion applies to a wide range of systems. 
For instance, the criterion works for the Liouvillian:
When the state space is given by a finite-dimensional Hilbert space $\mathcal{H}$, the Liouvillian acts on the Hilbert space $\mathcal{B}(\mathcal{H})$ spanned by all the linear operators on $\mathcal{H}$, which is also finite dimensional.
We can consider the $\Lambda$-localization for eigenstates of the Liouvillian on $\mathcal{B}(\mathcal{H})$ and discuss the corresponding skin effects.
Our criterion also works on skin effects without open boundaries \cite{Lee2021}.
Furthermore, our method applies to network models described by a linear operator, where no translation invariance usually exists.

%We hope to report these applications in the future.

%%%%% Acknowledgement %%%%%
We thank Daichi Nakamura, Yusuke O. Nakai, and Kohei Kawabata for their helpful discussions. 
This work was supported by JST CREST Grant No.~JPMJCR19T2, Japan.
K.S. was supported by the Graduate School of Science, Kyoto University, under the Ginpu Fund, and JST SPRING, Grant No.~JPMJSP2110.
M.S. was supported by JSPS KAKENHI Grant No.~JP24K00569.

%%%%% Note %%%%%
%\smallskip
\paragraph*{Note added.---} A part of this work was presented in \cite{Dresden}.
After completing this work, we became aware of a recent related work \cite{Shen2024}.

\bibliographystyle{apsrev4-2}
\bibliography{bib}

%%%%% Supplemental Material %%%%%
\newpage
\widetext

\renewcommand{\theequation}{S\arabic{equation}}
\renewcommand{\thefigure}{S\arabic{figure}}
\renewcommand{\thetable}{S\arabic{table}}
\renewcommand{\thesection}{S\arabic{section}}

\setcounter{equation}{0}
\setcounter{figure}{0}
\setcounter{table}{0}
\setcounter{section}{0}
\setcounter{Theorem}{0}

\begin{center}
{\bf \large Supplemental Material for %\\ \smallskip 
``\whatistitle"}
\end{center}

%%%%%%%%%%
\section{Proof of Theorem~\ref{thmSM:1}}\label{sec:proof_of_Fock_space_skin_effect}
In this section, we prove Theorem~\ref{thmSM:1}.
Below, we assume that the dimension $D$ of the Hilbert space $\mathcal{H}$ is finite and 
$\xi$ is an integer $\xi\in\bkb{1,2,\ldots,D-1}$.

\begin{Theorem}\label{thmSM:1}
    Let $\hat{H}$ be an operotor on $\mathcal{H}$, and $\ket{\psi_1},\ket{\psi_2},\ldots,\ket{\psi_{\xi+1}}\in\mathcal{H}$ be normalized eigenvectors of $\hat{H}$ associated with different eigenvalues.
    If all the $\xi+1$ eigenvectors $\ket{\psi_1},\ldots,\ket{\psi_{\xi+1}}$ are $\Lambda$-localized with the localization length $\xi$ for the threshold $\Lambda=\dfrac{1}{\sqrt{(\xi+1)(D-\xi)}}\eqqcolon\Lambda_\xi$, then $\hat{H}$ must be non-Hermitian.
\end{Theorem}

\begin{proof}
    We show that there is at least a nonorthogonal pair among the $\xi+1$ eigenvectors $\ket{\psi_1},\ldots,\ket{\psi_{\xi+1}}$ of $\hat{H}$, which leads to the non-Hermiticity of $\hat{H}$.

Let $(\ket{n})_{n=1}^D$ be a complete orthonormal basis of $\mathcal{H}$, and
suppose that $\ket{\psi_1},\ldots,\ket{\psi_{\xi+1}}$ are $\Lambda_\xi$-localized with the localization length $\xi$ with respect to the basis $(\ket{n})_n$.
Then, we introduce a $(\xi+1)\times(\xi+1)$ matrix $P$ whose elements are given by
\begin{align}
    (P)_{j,k}
    \coloneqq \braket{\psi_j|\psi_k}
    \quad
    (j,k=1,2,\ldots,\xi+1).
\end{align}
Note that every diagonal element of $P$ is one from the assumption.
Because we have
\begin{align}
    (P)_{j,k}
    = \sum_{n=1}^D \braket{\psi_j|n} \braket{n|\psi_k}
    = \sum_{n=1}^{\xi} \braket{n|\psi_j}^* \braket{n|\psi_k} + \sum_{n=\xi+1}^D \braket{n|\psi_j}^* \braket{n|\psi_k},
\end{align}
we can decompose $P$ into the sum of two $(\xi+1)\times(\xi+1)$ matrices $P_1$ and $P_2$ such that
\begin{gather}
    P
    = P_1 + P_2, \\
    P_1
    \coloneqq \begin{pmatrix}
        \braket{1|\psi_1} & \cdots & \braket{1|\psi_{\xi+1}} \\
        \vdots & \ddots & \vdots \\
        \braket{\xi|\psi_1} & \cdots & \braket{\xi|\psi_{\xi+1}}
    \end{pmatrix}^\dag
    \begin{pmatrix}
        \braket{1|\psi_1} & \cdots & \braket{1|\psi_{\xi+1}} \\
        \vdots & \ddots & \vdots \\
        \braket{\xi|\psi_1} & \cdots & \braket{\xi|\psi_{\xi+1}}
    \end{pmatrix}, \\
    P_2
    \coloneqq \begin{pmatrix}
        \braket{\xi+1|\psi_1} & \cdots & \braket{\xi+1|\psi_{\xi+1}} \\
        \vdots & \ddots & \vdots \\
        \braket{D|\psi_1} & \cdots & \braket{D|\psi_{\xi+1}}
    \end{pmatrix}^\dag
    \begin{pmatrix}
        \braket{\xi+1|\psi_1} & \cdots & \braket{\xi+1|\psi_{\xi+1}} \\
        \vdots & \ddots & \vdots \\
        \braket{D|\psi_1} & \cdots & \braket{D|\psi_{\xi+1}}
    \end{pmatrix}.
\end{gather}
Then, since the $(\xi+1)\times(\xi+1)$ matrix $P_1$ is a product of a $\xi\times (\xi+1)$ matrix and its Hermitian conjugation, it is positive semidefinite and of at most rank $\xi$.
Therefore, we can diagonalize $P_1$ by a unitary matrix $U$ as
\begin{align}
    U^\dag P_1 U
    = \diag(0,p_1,\ldots,p_\xi),
    \quad
    p_1,\ldots,p_\xi
    \ge 0.
\end{align}

Now, we prove the theorem by contradiction. For this purpose, we assume that $P$ is the identity matrix $I$.
Then, we have
\begin{align}
    U^\dag P_2 U 
    = I - \diag(0,p_1,\ldots,p_\xi)
    = \diag(1,1-p_1,\ldots,1-p_\xi),
\end{align}
which requires that $P_2$ has the eigenvalue $1$.
However, from the  assumption that $\ket{\psi_1},\ldots,\ket{\psi_{\xi+1}}$ are $\Lambda_\xi$-localized with the localization length $\xi$,
we can show that the spectral radius $\rho(P_2)$ of $P_2$ is smaller than 1:
\begin{align}
    \rho(P_2)
    \le \tr P_2
    = \sum_{j=1}^{\xi+1}\sum_{n=\xi+1}^D\abs{\braket{n|\psi_j}}^2
    < \sum_{j=1}^{\xi+1}\sum_{n=\xi+1}^D\bka{\frac{1}{\sqrt{(\xi+1)(D-\xi)}}}^2
    = 1,
\end{align}
which implies that $P_2$ cannot have the eigenvalue $1$. As a result,
we conclude that $P$ is not the identity matrix, which indicates there exists a nonzero component of $(P)_{j,k} = \braket{\psi_j|\psi_k}$ ($j\neq k$).
Because eigenvectors with different eigenvalues must be orthogonal for a Hermitian (more precisely normal) Hamiltonian, $\hat{H}$ must be non-Hermitian (non-normal).
\end{proof}

\section{Remarks on the tightness of Theorem~\ref{thmSM:1}}\label{sec:tightness}

In this section, we discuss how tight the evaluation in the Theorem~\ref{thmSM:1} is.

Theorem~\ref{thmSM:1} describes a trade-off between the localization length $\xi$ and the number of $\Lambda_\xi$-localized eigenvectors in Hermitian systems: Hermitian systems do not allow more than $\xi$ $\Lambda_\xi$-localized eigenvectors within the localization length $\xi$.
We can also show that Hermitian systems allow less than $\xi+1$ $\Lambda_\xi$-localized eigenvectors within the localization length $\xi$:
Since eigenvectors in Hermitian systems are orthonormal, we can choose the eigenvectors as the basis to define the $\Lambda_\xi$-localization. Then, we have less than $\xi+1$ $\Lambda_\xi$-localized eigenvectors within the localization length $\xi$.
Therefore, $\xi+1$ is the minimal number of $\Lambda_\xi$-localized eigenvectors required in Theorem~\ref{thmSM:1}.

Next, we discuss how tight the threthold $\Lambda_\xi=\dfrac{1}{\sqrt{(\xi+1)(D-\xi)}}$ in Theorem~\ref{thmSM:1} is.
For this purpose, we introduce $\xi+1$ vectors $\ket{\psi_1},\ldots,\ket{\psi_{\xi+1}}$ for the basis $(\ket{n})_n$ as
%\begin{align}
%    \ket{\psi_j}
%    = \transpose{\Bigg(\underbrace{\dfrac{e^{2\pi in/(\xi+1)}}{\sqrt{\xi+1}} \ \dfrac{e^{4\pi in/(\xi+1)}}%{\sqrt{\xi+1}} \ \cdots \ \dfrac{e^{2\xi\pi in/(\xi+1)}}{\sqrt{\xi+1}}}_{\xi} \ \underbrace{\dfrac{1}{\sqrt{(\xi+1)(D-\xi)}} \ \cdots \ \dfrac{1}{\sqrt{(\xi+1)(D-\xi)}}}_{D-\xi}\Bigg)},
%\end{align}
\begin{align}
    \ket{\psi_j}
    = \underbrace{\dfrac{e^{2\pi ij/(\xi+1)}}{\sqrt{\xi+1}}\ket{1}+\dfrac{e^{4\pi ij/(\xi+1)}}{\sqrt{\xi+1}}\ket{2}+\cdots+\dfrac{e^{2\xi\pi ij/(\xi+1)}}{\sqrt{\xi+1}}\ket{\xi}}_{\xi}
    + \underbrace{\dfrac{1}{\sqrt{(\xi+1)(D-\xi)}}\ket{\xi+1} +\cdots +\dfrac{1}{\sqrt{(\xi+1)(D-\xi)}}\ket{D}}_{D-\xi}
\end{align}
which are $\Lambda$-localized with the localization length $\xi$ for the threshold $\Lambda>\Lambda_\xi$.
Then, $\ket{\psi_1},\ldots,\ket{\psi_{\xi+1}}$ are found to be orthonormal each other,
\begin{align}
    \braket{\psi_j|\psi_k}
    = \delta_{j,k},
\end{align}
so we can regard them as eigenstates of a Hermitian Hamiltonian.
Therefore, we can put $\xi+1$ $\Lambda$-localized states inside the localization length $\xi$ for a Hermitian Hamiltonian once we choose the threshold $\Lambda$ larger than $\Lambda_\xi$.
Hence, $\Lambda$ must be less than or equal to $\Lambda_\xi$ in Theorem~\ref{thmSM:1}.

\section{Localization criterion for the Hatano-Nelson model}\label{sec:HatanoNelson}
\begin{figure}[htbp]
    \centering
    \includegraphics[width=\columnwidth,clip]{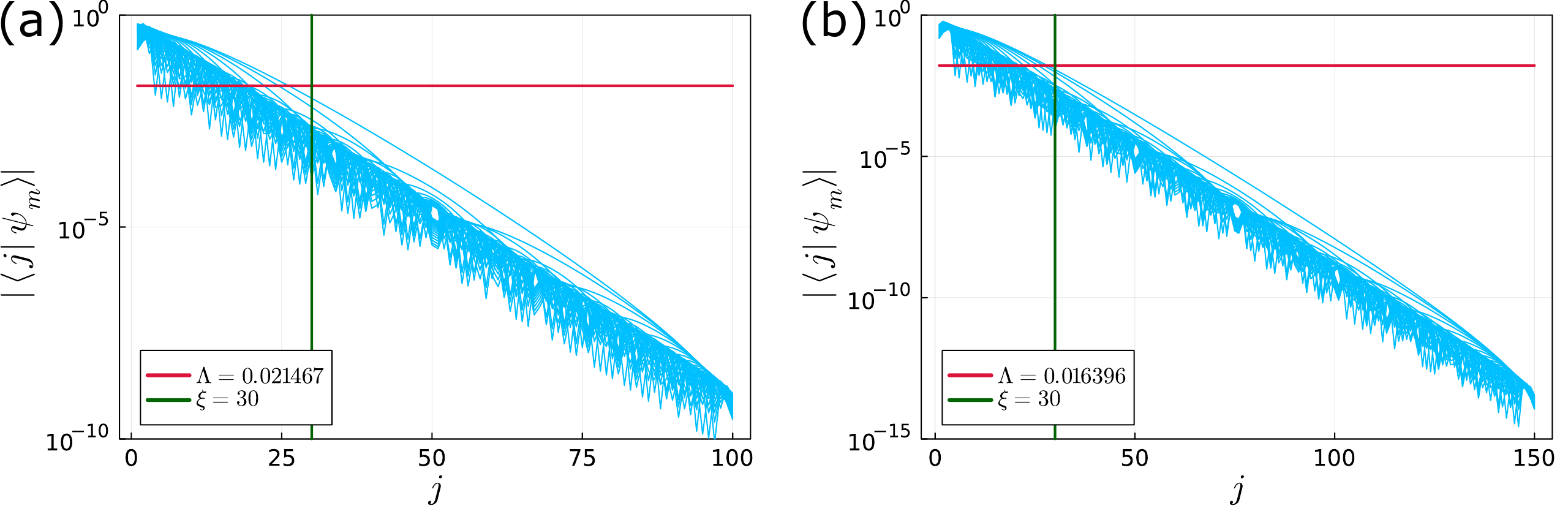}
    \caption{Verification of the localization criterion for the single-particle Hatano-Nelson model ($t=1.0$, $g=0.2$). 
    We take the basis $\ket{j}$ in real space coordinate and the localization length $\xi=30$.
    (a) For $L=100$, all of the amplitudes $\abs{\braket{j|\Psi_m}}$ are smaller than the threshold $\Lambda_\xi=0.021\,467$ (the red solid line) for $n\ge\xi+1$, with $D=L$.
    Thus, the Hatano-Nelson model satisfies the general criterion of non-Hermitian skin effects in Theorem~\ref{thmSM:2}.
    (b) For $L=150$, one can also confirm the general criterion, though the threshold $\Lambda_\xi$ becomes $0.016\,396$ due to the change of the dimension $D=L$.}
    \label{Sfig:HatanoNelson_LambdaLoc}
\end{figure}

In this section, we examine the single-particle Hatano-Nelson model (without disorders)
\begin{align}
    \hat{H}
    = \sum_{j=1}^{L-1}\bka{(t+g)\ket{j+1}\bra{j} + (t-g)\ket{j}\bra{j+1}},
\end{align}
from the viewpoint of our general criterion for non-Hermitian skin effects in Theorem~\ref{thmSM:1}.
In Fig. \ref{Sfig:HatanoNelson_LambdaLoc}, we illustrate the non-Hermitian skin effect of the single-particle Hatano-Nelson model. As is seen clearly, the single-particle Hatano-Nelson model satisfies our general criterion for non-Hermitian skin effects.

%\section{Definition and properties of the 
%condition number}\label{sec:condition_number}

% \section{Algorithm for verifying the Fock space skin effect of the fermionic Hatano-Nelson model with interaction}\label{sec:algorithm}

\section{Proof of Theorem~\ref{thmSM:2}}\label{sec:proof_of_condition_number_inequality}

In this section, we prove Theorem~\ref{thmSM:2}.
The accurate statement of Theorem~\ref{thmSM:2} is as follows.

\begin{Theorem}\label{thmSM:2}
    Let $\hat{H}$ be a diagonalizable operator on a finite-dimensional Hilbert space $\mathcal{H}$ of the dimension $D$, and $\ket{\psi_1},\ldots,\ket{\psi_D}$ be linearly independent normalized eigenvectors of $\hat{H}$.
    For an orthonormal basis $(\ket{n})_{n=1}^D$ of $\mathcal{H}$,
    suppose that $\ket{\psi_1},\ldots,\ket{\psi_\xi},\ket{\psi_{\xi+1}}$ are $\Lambda$-localized with the localization length $\xi\in\bkb{1,\ldots,D-1}$ for the threshold $\Lambda\le\dfrac{1}{\sqrt{D-\xi}}$. 
    Then, the condition number $\kappa(\hat{V})$ of the operator $\hat{V}\coloneqq\sum_{n=1}^D\ket{\psi_n}\bra{n}$ diagonalizing $\hat{H}$ satisfies
    \begin{align}
        \kappa(\hat{V})
        > \sqrt{\frac{1}{(\xi+1)(D-\xi)\Lambda^2}-\frac{1}{\xi+1}}
        = \sqrt{\frac{\Lambda_\xi^2}{\Lambda^2}-\frac{1}{\xi+1}}.
    \end{align}
\end{Theorem}

%We have already defined the condition number for diagonalizable matrices in the main text or Sec.~\ref{sec:condition_number}.
%Note that the definition does not depend on the choice of the orthonormal basis.

To prove this theorem,  we first generalize the condition number to the case where the matrix is not square.

\begin{Definition}
    We define the condition number $\kappa(A)$ of an $m\times n$ matrix $A$ as
\begin{align}
    \kappa(A)
    \coloneqq \frac{\sigma_{\mathrm{max}}(A)}{\sigma_{\mathrm{min}}(A)},
\end{align}
where $\sigma_{\mathrm{max}}(A)$ is the maximum singular value of $A$ and $\sigma_{\mathrm{min}}(A)$ is the minimum one.
For $\sigma_{\mathrm{min}}(A) = 0$, we formally define $\kappa(A)\coloneqq +\infty$.
\end{Definition}

This definition includes the original definition for square matrices.
For singular values, we have the following property:
\begin{Proposition}[For the proof, see Corollary 3.1.3 in Ref.\cite{Horn1991}.]\label{prop:singular_value}
    Let $A$ be an $m\times n$ matrix, $A_r$ be an $m\times(n-r)$ submatrix of $A$ obtained by deleting $r$ columns from $A$, and $\sigma_k(X)$ be the $k$-th largest singular value of a $p\times q$ matrix $X$:
    \begin{align}
        \sigma_{\mathrm{max}}(X)
        = \sigma_1(X)
        \ge \cdots
        \ge \sigma_{\min\bkb{p,q}}
        = \sigma_{\mathrm{min}}(X)
        \ge 0.
    \end{align}
    Then, it holds that
    \begin{align}
        \sigma_k(A)
        \ge \sigma_k(A_r)
        \ge \sigma_{k+r}(A),
        \quad
        k
        = 1,\ldots,\min\bkb{m,n},
    \end{align}
    where for a $p\times q$ matrix $X$ we set $\sigma_j(X)\coloneqq 0$ if $j>\min\bkb{p,q}$.
\end{Proposition}
From this Proposition, we immediately obtain the following Proposition.
\begin{Proposition}\label{prop:condition_number}
    Let $A$ be a $D\times D$ matrix, and $A_r$ be a $D\times(D-r)$ submatrix
    of $A$ obtained by deleting $r$ columns from $A$.
    Then, it holds that
    \begin{align}
        \kappa(A_r)
        \le \kappa(A).
    \end{align}
\end{Proposition}
%\begin{proof}[Proof of Proposition~\ref{prop:condition_number}]
%    The $D\times(D-r)$ matrix $A_r$ is obtained by deleting $r$ columns from $A$, so that the inequalities
%    \begin{align}
%        \sigma_1(A)
%        \ge \sigma_1(A_r),
%        \quad
%        \sigma_{D-r}(A_r)
%        \ge \sigma_D(A)
%    \end{align}
%    hold.
%    Thus, we have
%    \begin{align}
%        \kappa(A_r)
%        = \frac{\sigma_1(A_r)}{\sigma_{D-r}(A_r)}
%        \le \frac{\sigma_1(A)}{\sigma_D(A)}
%        = \kappa(A).
%    \end{align}
%\end{proof}

We also have Proposition~\ref{prop:eigenvalue}.

\begin{Proposition}[For proof, see Theorem 4.3.7 in Ref.\cite{Horn2012}.] \label{prop:eigenvalue}
    Let $A,B$ be $n\times n$ Hermitian matrices.
    Arrange the eigenvalues of $A$ and $B$ in the following way:
    \begin{align}
        \lambda_1(A)
        \le \lambda_2(A)
        \le \cdots
        \le \lambda_n(A),
        \quad
        \lambda_1(B)
        \le \lambda_2(B)
        \le \cdots
        \le \lambda_n(B).
    \end{align}
    Then, we have
    \begin{align}
        \lambda_j(A) + \lambda_1(B)
        \le \lambda_j(A+B)
        \le \lambda_j(A) + \lambda_n(B),
        \quad
        j
        = 1,\ldots,n.
    \end{align}
\end{Proposition}

Now we prove Theorem~\ref{thmSM:2}.

\begin{proof}[Proof of Theorem~\ref{thmSM:2}]
    Let $V$ be the $D\times D$ matrix whose elements are given by
    \begin{align}
        (V)_{m,n}
        \coloneqq \bra{m}\hat{V}\ket{n}
        = \braket{m|\psi_n},
        \quad
        m,n
        = 1,2,\ldots,D
    \end{align}
    and $V'$ be the $D\times(\xi+1)$ submatrix of $V$ whose elements given by
    \begin{align}
        (V')_{m,n}
        \coloneqq \braket{m|\psi_n},
        \quad
        m
        = 1,2,\ldots,D,
        \quad
        n
        = 1,\ldots,\xi+1.
    \end{align}
    Since $V'$ is obtained by deleting $D-\xi-1$ columns of $V$, we have
    \begin{align}
        \kappa(\hat{V})
        = \kappa(V)
        \ge \kappa(V')
    \end{align}
    from Proposition~\ref{prop:condition_number}.
    To prove Theorem~\ref{thmSM:2}, we will show that
    \begin{align}
        \kappa(V')
        > \sqrt{\frac{1}{(\xi+1)(D-\xi)\Lambda^2}-\frac{1}{\xi+1}}.
    \end{align}
    For this purpose, we introduce the $(\xi+1)\times(\xi+1)$ positive-semidefinite Hermitian matrix,
    \begin{align}
        P
        \coloneqq (V')^\dag V'.
    \end{align}
   Then, we obtain
    \begin{align}
        \kappa(V')
        = \frac{\sigma_{\mathrm{max}}(V')}{\sigma_{\mathrm{min}}(V')}
        = \sqrt{\frac{\lambda_{\mathrm{max}}(P)}{\lambda_{\mathrm{min}}(P)}},
    \end{align}
 where $\lambda_{\mathrm{max}}(P)$ and $\lambda_{\mathrm{min}}(P)$ are the maximum and minimum eigenvalue of $P$, respectively.
 Now we decompose $P$ into the sum of two $(\xi+1)\times(\xi+1)$ matrices $P_1$ and $P_2$ in the same way as the proof of Theorem~\ref{thmSM:1}:
    \begin{gather}
        P
        = P_1 + P_2, \\
        P_1
        \coloneqq \begin{pmatrix}
            \braket{1|\psi_1} & \cdots & \braket{1|\psi_{\xi+1}} \\
            \vdots & \ddots & \vdots \\
            \braket{\xi|\psi_1} & \cdots & \braket{\xi|\psi_{\xi+1}}
        \end{pmatrix}^\dag
        \begin{pmatrix}
            \braket{1|\psi_1} & \cdots & \braket{1|\psi_{\xi+1}} \\
            \vdots & \ddots & \vdots \\
            \braket{\xi|\psi_1} & \cdots & \braket{\xi|\psi_{\xi+1}}
        \end{pmatrix}, \\
        P_2
        \coloneqq \begin{pmatrix}
            \braket{\xi+1|\psi_1} & \cdots & \braket{\xi+1|\psi_{\xi+1}} \\
            \vdots & \ddots & \vdots \\
            \braket{D|\psi_1} & \cdots & \braket{D|\psi_{\xi+1}}
        \end{pmatrix}^\dag
        \begin{pmatrix}
            \braket{\xi+1|\psi_1} & \cdots & \braket{\xi+1|\psi_{\xi+1}} \\
            \vdots & \ddots & \vdots \\
            \braket{D|\psi_1} & \cdots & \braket{D|\psi_{\xi+1}}
        \end{pmatrix}.
    \end{gather}
    Both of $P_1$ and $P_2$ are $(\xi+1)\times(\xi+1)$ positive-semidefinite Hermitian matrices, and $P_1$ is of at most rank $\xi$, so the minimum eigenvalue of $P_1$ becomes zero:
    \begin{align}
        \lambda_{\mathrm{min}}(P_1)
        = 0.
    \end{align}
    From Proposition~\ref{prop:eigenvalue}, we have
    \begin{gather}
        \lambda_{\mathrm{max}}(P)
        \ge \lambda_{\mathrm{max}}(P_1) + \lambda_{\mathrm{min}}(P_2)
        \ge \lambda_{\mathrm{max}}(P_1), \\
        \lambda_{\mathrm{min}}(P)
        \le \lambda_{\mathrm{min}}(P_1) + \lambda_{\mathrm{max}}(P_2)
        = \lambda_{\mathrm{max}}(P_2),
    \end{gather}
    which leads to
    \begin{align}
        \kappa(V')
        \ge \sqrt{\frac{\lambda_{\mathrm{max}}(P_1)}{\lambda_{\mathrm{max}}(P_2)}}.
    \end{align}

    Now we evaluate $\lambda_{\mathrm{max}}(P_1)$ from below and $\lambda_{\mathrm{max}}(P_2)$ from above by employing the $\Lambda$-localization property of $\ket{\psi_1},\ldots,\ket{\psi_{\xi+1}}$.
    For $\lambda_{\mathrm{max}}(P_1)$, we have
    \begin{align}
        \lambda_{\mathrm{max}}(P_1)
        &= \sigma_{\mathrm{max}}\bka{\begin{pmatrix}
            \braket{1|\psi_1} & \cdots & \braket{1|\psi_{\xi+1}} \\
            \vdots & \ddots & \vdots \\
            \braket{\xi|\psi_1} & \cdots & \braket{\xi|\psi_{\xi+1}}
        \end{pmatrix}}^2
        = \norm{\begin{pmatrix}
            \braket{1|\psi_1} & \cdots & \braket{1|\psi_{\xi+1}} \\
            \vdots & \ddots & \vdots \\
            \braket{\xi|\psi_1} & \cdots & \braket{\xi|\psi_{\xi+1}}
        \end{pmatrix}}^2 \\
        &\ge \norm{\begin{pmatrix}
            \braket{1|\psi_1} & \cdots & \braket{1|\psi_{\xi+1}} \\
            \vdots & \ddots & \vdots \\
            \braket{\xi|\psi_1} & \cdots & \braket{\xi|\psi_{\xi+1}}
        \end{pmatrix}\begin{pmatrix}
            1 \\ 0 \\ \vdots \\ 0
        \end{pmatrix}}^2
        = \sum_{n=1}^\xi\abs{\braket{n|\psi_1}}^2
        = 1-\sum_{n=\xi+1}^D\abs{\braket{n|\psi_1}}^2 \\
        &> 1-(D-\xi)\Lambda^2.
    \end{align}
    For $\lambda_{\mathrm{max}}(P_2)$, we have
    \begin{align}
        \lambda_{\mathrm{max}}(P_2)
        \le \tr P_2
        = \sum_{n=\xi+1}^D\sum_{j=1}^{\xi+1}\abs{\braket{n|\psi_j}}^2
        < (\xi+1)(D-\xi)\Lambda^2,
    \end{align}
because any eigenvalue of $P_2$ is zero or positive. Thus, we eventually obtain
    \begin{align}
        \kappa(\hat{V})
        \ge \kappa(V')
        > \sqrt{\frac{1-(D-\xi)\Lambda^2}{(\xi+1)(D-\xi)\Lambda^2}}
        = \sqrt{\frac{1}{(\xi+1)(D-\xi)\Lambda^2} - \frac{1}{\xi+1}}.
    \end{align}
\end{proof}

\section{Spectra of interacting fermionic Hatano-Nelson model under the periodic boundary condition and the open boundary condition}
\label{sec:spectra}

\begin{figure}
    \centering
    \includegraphics[width=0.5\linewidth,clip]{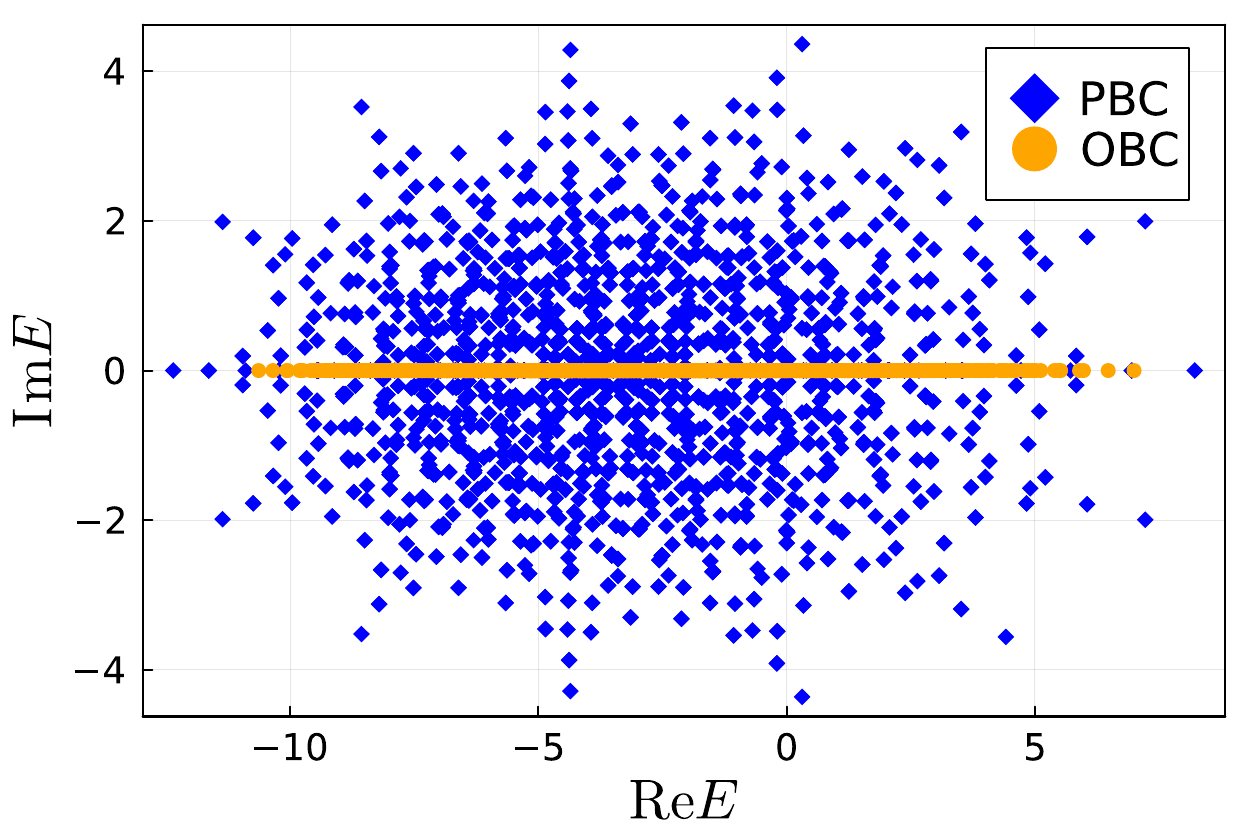}
    \caption{The spectra of the fermionic Hatano-Nelson model with interaction under the periodic boundary condition (PBC) and the open boundary condition (OBC).
    The model parameters are $\alpha=0.5$, $U=-1$, $L=14$, and $N=L/2$ (half-filling).
    The two spectra are entirely different.}
    \label{Sfig:spectra_L14}
\end{figure}

The fermionic Hatano-Nelson model with interaction exhibits the spectrum change between the periodic and the open boundary conditions \cite{Kawabata2022}.
Figure \ref{Sfig:spectra_L14} shows the spectra of the fermionic Hatano-Nelson model with interaction under the periodic and open boundary conditions: While the periodic boundary condition gives a complex spectrum, the open boundary condition gives a real spectrum. This difference can be understood by a similarity transformation called the imaginary gauge transformation. The similarity transformation is given by
\begin{align}
\hat{R}=\prod_{j=1}^L(1+(e^{-\alpha j}-1)\hat{n}_j),    
\label{eq:similarity_trans}
\end{align}
which maps the interacting non-Hermitian Hatano-Nelson model to a Hermitian one under the open boundary condition \cite{Hamazaki2019},
\begin{align}
\hat{R}^{-1}\hat{H}_{\rm iHN}\hat{R}=\sum_{j=1}^{L-1}\left(\hat{c}_j^\dagger \hat{c}_{j+1}+\hat{c}_{j+1}^\dagger \hat{c}_j+U\hat{n}_j\hat{n}_{j+1}
\right).
\quad 
\label{eq:similarity}
\end{align}
Since the similarity transformation does not change the spectrum, we can conclude that the spectrum of the interacting Hatano-Nelson model coincides with that of the Hermitian model, and thus it is real.
We also note that Eq.(\ref{eq:similarity}) does not hold for the periodic boundary condition: The boundary terms connecting site $L$ and site 1 remain non-Hermitian even after the similarity transformation.
Thus, the periodic boundary condition gives the interacting Hatano-Nelson model a complex spectrum.

\section{Estimation of $\kappa_0$ for the interacting fermionic Hatano-Nelson model}\label{sec:derive}
Here, we estimate $\kappa_0$ for the interacting fermionic Hatano-Nelson model. 
For this purpose, we use the similarity transformation in Eq.(\ref{eq:similarity}).
Using the similarity transformation,
we can get the eigenstates $\ket{\Psi_m}$ of the interacting Hatano-Nelson model from the eigenstates $\ket{\Phi_m}$ of the Hermitian model on the right-hand side in Eq.(\ref{eq:similarity}),
\begin{align}
\ket{\Psi_m}=\hat{R}\ket{\Phi_m}/\|\hat{R}\ket{\Phi_m}\|.    
\end{align}
where we have normalized $\ket{\Psi_m}$ as $\braket{\Psi_m|\Psi_m}=1$.
Then, we have
 \begin{align}
\langle n|\Psi_m\rangle&=\langle n|\hat{R}\ket{\Phi_m}    
/\|\hat{R}\ket{\Phi_m}\|   
\nonumber\\
&=\frac{\sum_{p}\bra{n}\hat{R}\ket{p}\braket{p|\Phi_m}}{
\|\sum_{q} \hat{R}\ket{q}\braket{q|\Phi_m}\|}
\nonumber\\
&=\frac{\bra{n}\hat{R}\ket{n}\braket{n|\Phi_m}}{
\|\sum_{q} \hat{R}\ket{q}\braket{q|\Phi_m}\|},
\label{eq:cefficient}
\end{align}
where $(\ket{n})_n$ is the Fock basis with the particle number $N$ and we have used the fact that $\ket{n}$ is an eigenstate of $\hat{R}$.
As mentioned in the main text, we arrange the Fock basis as
\begin{align}
|\braket{1|\Psi_1}|\ge |\braket{2|\Psi_1}|\ge \dots \ge|\braket{\dim {\cal F}_N|\Psi_1}|,    
\end{align}
where we have chosen $\Psi_1$ as the reference state.
Assuming that no particular localization occurs in the eigenstates $\Phi_m$ of the Hermitian Hamiltonian, 
the coefficient $\braket{n|\hat{R}|n}$ in  Eq.(\ref{eq:cefficient}) determines the order of the Fock basis.
Then, noting that 
\begin{align}
\max_{n}\braket{n|\hat{R}|n}=\exp\left(-\alpha\sum_{j=1}^Nj\right), 
\quad
\min_{n}\braket{n|\hat{R}|n}=\exp\left(-\alpha\sum_{j=L-N+1}^Lj\right), \quad (\alpha>0)
\end{align}
we can evaluate $|\braket{\dim {\cal F}_N|\Psi_m}|$ as
\begin{align}
|\braket{\dim {\cal F}_N|\Psi_m}|
&=\frac{|\bra{\dim {\cal F}_N}\hat{R}\ket{\dim {\cal F}_N}\braket{\dim {\cal F}_N|\Phi_m}|}
{\|\sum_{q} \hat{R}\ket{q}\braket{q|\Phi_m}\|}
\nonumber\\
&\sim \frac{\exp\bka{-\alpha\sum_{j=L-N+1}^{L} j}}{\|\hat{R}\ket{1}\|}
\nonumber\\
&=\frac{\exp\bka{-\alpha\sum_{j=L-N+1}^{L} j}}{\exp\bka{-\alpha\sum_{j=1}^N j}}=\exp(\alpha N(L-N)), 
\end{align}
for $\alpha>0$ and $L\gg 1$.
As a result, we have the estimation of $\kappa_0$ in the main text:
\begin{align}
\kappa_0\sim \exp(\alpha N(L-N))/\sqrt{\dim {\cal F}_N}.    
\end{align}
%Lastly, let us remark on the relation between Theorem~\ref{thmSM:1} and \ref{thmSM:2}.
%If we take the threshold $\Lambda$ as $\Lambda\le\dfrac{1}{\sqrt{(\xi+2)(D-\xi)}}$, Theorem~\ref{thmSM:2} implies that $\kappa(\hat{V}) > 1$ and thus $\hat{H}$ must be non-Hermitian.
%This implication is similar to Theorem~\ref{thmSM:1}, but does not include Theorem~\ref{thmSM:1}, because the choice $\Lambda\le\dfrac{1}{\sqrt{(\xi+2)(D-\xi)}}$ of threshold does not cover $\Lambda_\xi=\dfrac{1}{\sqrt{(\xi+1)(D-\xi)}}$ in Theorem~\ref{thmSM:1}.
%As for judging non-Hermiticity of a given Hamiltonian, Theorem~\ref{thmSM:1} is stricter than Theorem~\ref{thmSM:2} properly.
%Indeed, we prove Theorem~\ref{thmSM:1} and \ref{thmSM:2} independently each other.
%One might improve the proof of Theorem~\ref{thmSM:2} to obtain a tighter lower bound, but that would not be straightforward.

\section{Evolution of survival probability}\label{sec:prob}
In this section, we show that the survival probability in the main text satisfies
\begin{align}\label{eq:surviving_prob}
    P(t;\hat{\rho}_{\mathrm{init}})
    = \tr\bka{e^{-i\hat{H}_{\mathrm{eff}}t}\hat{\rho}_{\mathrm{init}}e^{i\hat{H}_{\mathrm{eff}}^\dag t}}.
\end{align}

We derive this equation from the Lindblad equation,
\begin{align}\label{Seq:Lindblad0}
    \frac{\partial\hat{\rho}}{\partial t}
    &= -i(\hat{H}_0\hat{\rho}-\hat{\rho}\hat{H}_0) + \sum_r\left(\hat{L}_r\hat{\rho}\hat{L}_r^\dag - \frac{1}{2}\{\hat{L}_r^\dag\hat{L}_r,\hat{\rho}\}\right) \nonumber\\
    &= -i(\hat{H}_{\mathrm{eff}}\hat{\rho}-\hat{\rho}\hat{H}_{\mathrm{eff}}^\dag) + \sum_r \hat{L}_r\hat{\rho}\hat{L}_r^\dag. 
\end{align}
with the effective Hamiltonian $\hat{H}_{\rm eff}$ is defined by 
\begin{align}\label{Seq:effectiveHam}
    \hat{H}_{\mathrm{eff}}
    = \hat{H}_0 - \frac{i}{2}\sum_{r}\hat{L}_r^\dag\hat{L}_r. 
\end{align}
We impose the following two conditions on the Lindblad equation:
\begin{itemize}
    \item The Hermitian Hamiltonian $\hat{H}_0$ preserves the particle number $\hat{N}$:
    \begin{align}
        [\hat{H}_0,\hat{N}]
        = 0,
        \quad
        \hat{N}
        = \sum_{j=1}^L \hat{c}_j^\dag\hat{c}_j.
    \end{align}
    \item The Lindblad operator $\hat{L}_r$ gives an $n_r$-body loss $(n_r\neq 0)$:
    \begin{align}
        \hat{L}_r
        = \sum_{j_1,\ldots,j_{n_r}} d_{j_1,\ldots,j_{n_r}}\prod_{\alpha=1}^{n_r}\hat{c}_{j_\alpha}
        \quad
        \bka{d_{j_1,\ldots,j_{n_r}}\in\C}.
    \end{align}
\end{itemize}
The operators $\hat{H}_0$ and $\hat{L}_r$ in the main text satisfy these conditions.
Let $\hat{P}_N$ be the orthogonal projection onto the $N$-particle Fock space $\mathcal{F}_N$.
For $N>L$, we define $\hat{P}_N\coloneqq 0$ for convenience.
Since we have
\begin{align}
    \hat{P}_N\hat{P}_{N'}
    = \delta_{N,N'}\hat{P}_N,
    \quad
    \sum_N \hat{P}_N
    = \hat{I},
\end{align}
any operator $\hat{A}$ can be recast into
\begin{align}
    \hat{A}
    = \pmat{
        \hat{P}_0 & \hat{P}_1 & \hat{P}_2 & \cdots
    }\pmat{
        \hat{A}_{0,0} & \hat{A}_{0,1} & \hat{A}_{0,2} & \cdots \\
        \hat{A}_{1,0} & \hat{A}_{1,1} & \hat{A}_{1,2} & \cdots \\
        \hat{A}_{2,0} & \hat{A}_{2,1} & \hat{A}_{2,2} & \cdots \\
        \vdots & \vdots & \vdots & \ddots
    }\pmat{
        \hat{P}_0 \\ \hat{P}_1 \\ \hat{P}_2 \\ \vdots
    },
    \quad
    \hat{A}_{m,n}
    \coloneqq \hat{P}_m\hat{A}\hat{P}_n.
\end{align}

If the initial state $\hat{\rho}_{\mathrm{init}}$ consists of $N$-particles, we have $\hat{\rho}_{\rm init} = \hat{P}_N\hat{\rho}_{\rm init}\hat{P}_N$. Then, the survival probability is of the form
\begin{align}
    P(t;\hat{\rho}_{\mathrm{init}})
    = \tr(\hat{P}_N \hat{\rho}(t) \hat{P}_N)
    = \tr(\hat{\rho}_{N,N}(t)),
\end{align}
from the definition.
Now, we consider the time evolution of  $\hat{\rho}_{m,n}=\hat{P}_m\hat{\rho}\hat{P}_n$.
Since we have
\begin{align}
    \hat{P}_N\hat{c}_j
    = \hat{c}_j\hat{P}_{N+1}, 
    \quad
    \hat{c}_j^\dag\hat{P}_N
    = \hat{P}_{N+1}\hat{c}_j^\dag
\end{align}
and thus
\begin{align}
    [\hat{H}_{\rm eff},\hat{P}_N]
    = 0,
    \quad
    \hat{P}_N\hat{L}_r
    = \hat{L}_r\hat{P}_{N+n_r},
    \quad
    \hat{L}_r^\dag\hat{P}_N
    = \hat{P}_{N+n_r}\hat{L}_r^\dag,
\end{align}
it holds that
\begin{align}\label{eq:time_evolution_rhomn}
    \frac{\partial\hat{\rho}_{m,n}}{\partial t}
    &= -i\hat{P}_m\bka{\hat{H}_{\rm eff}\hat{\rho}-\hat{\rho}\hat{H}_{\rm eff}^\dag}\hat{P}_n + \sum_r \hat{P}_m\hat{L}_r\hat{\rho}\hat{L}_r^\dag\hat{P}_n \nonumber\\
    &= -i\bka{\hat{H}_{\rm eff}\hat{\rho}_{m,n}-\hat{\rho}_{m,n}\hat{H}_{\rm eff}^\dag} + \sum_r\hat{L}_r\hat{\rho}_{m+n_r,n+n_r}\hat{L}_r^\dag.
\end{align}
Then, by introducing
\begin{align}
    \hat{\rho}'(t)
    \coloneqq e^{i\hat{H}_{\rm eff} t}\hat{\rho}(t)e^{-i\hat{H}_{\rm eff}^\dag t},
\end{align}
the time evolution equation \eqref{eq:time_evolution_rhomn} reads
\begin{align}
    \frac{\partial\hat{\rho}'_{m,n}}{\partial t}
    &= e^{i\hat{H}_{\rm eff} t}\bka{\sum_r\hat{L}_r\hat{\rho}_{m+n_r,n+n_r}(t)\hat{L}_r^\dag}e^{-i\hat{H}_{\rm eff}^\dag t} \\
    &= \sum_r \hat{L}_r'(t)\hat{\rho}'_{m+n_r,n+n_r}(t)\hat{L}_r'{}(t)^\dag,
\end{align}
with $\hat{L}_r'(t) \coloneqq e^{i\hat{H}_{\rm eff} t}\hat{L}_r e^{-i\hat{H}_{\rm eff}^\dag t}$.
The solution of this equation is given by
\begin{align}
    \hat{\rho}'_{m,n}(t_0)
    &= \hat{\rho}_{m,n}(0) + \int_0^{t_0} \dif t_1 \sum_{r_1}\hat{L}'_{r_1}(t_1)\hat{\rho}_{m+n_{r_1},n+n_{r_1}}'(t_1)\hat{L}_{r_1}'(t_1)^\dag \\
    &\ \begin{aligned}
        =\ \hat{\rho}_{m,n}(0)\ +\  
        &\sum_{k=1}^\infty \sum_{r_1,\ldots r_k}\int_0^{t_0}\dif t_1\cdots \int_0^{t_{k-1}}\dif t_k \\
        &\times\bka{\hat{L}'_{r_1}(t_1)\cdots\hat{L}'_{r_k}(t_k)}\hat{\rho}_{m+n_{r_1}+\cdots+n_{r_k},n+n_{r_1}+\cdots+n_{r_k}}(0)\bka{\hat{L}'_{r_1}(t_1)\cdots\hat{L}'_{r_k}(t_k)}^\dag.
    \end{aligned}
\end{align}
If we take $\hat{\rho}(0) = \hat{\rho}_{\rm init} = \hat{P}_N\hat{\rho}_{\rm init}\hat{P}_N$ and $m=n=N$, $\hat{\rho}_{m+n_{r_1}+\cdots+n_{r_k},n+n_{r_1}+\cdots+n_{r_k}}(0)$ in the second term of the above equation vanishes. Thus, we obtain
\begin{align}
    \hat{\rho}'_{N,N}(t)
    = \hat{\rho}_{N,N}(0),
    \quad
    {\it i.e.},
    \quad
    \hat{\rho}_{N,N}(t)
    = e^{-i\hat{H}_{\rm eff} t}\hat{\rho}_{N,N}(0)e^{i\hat{H}_{\rm eff}^\dag t}
    = e^{-i\hat{H}_{\rm eff} t}\hat{\rho}_{\rm init}e^{i\hat{H}_{\rm eff}^\dag t},
\end{align}
which leads to Eq.\eqref{eq:surviving_prob}.

Note that whereas the above derivation employs the full Lindblad equation in Eq. (\ref{Seq:Lindblad0}), only the effective Hamiltonian $\hat{H}_{\rm eff}$ governs the survival probability. 
The quantum jump term $\sum_{r}\hat{L}_r\hat{\rho}\hat{L}_r^\dagger$ in Eq.(\ref{Seq:Lindblad0}) is irrelevant to the survival probability since it changes the total particle number.

\section{Experimental feasibility of Fock space skin effect in ultracold fermionic gases}
In this section, we discuss the experimental feasibility of the Fock space skin effect in ultracold fermionic gases on an optical lattice.

In the main text, we consider an open quantum system governed by the Lindblad equation
\begin{align}\label{Seq:Lindblad}
    \frac{\partial\hat{\rho}}{\partial t}
    &= -i(\hat{H}_0\hat{\rho}-\hat{\rho}\hat{H}_0) + \sum_r\left(\hat{L}_r\hat{\rho}\hat{L}_r^\dag - \frac{1}{2}\{\hat{L}_r^\dag\hat{L}_r,\hat{\rho}\}\right) \nonumber\\
    &= -i(\hat{H}_{\mathrm{eff}}\hat{\rho}-\hat{\rho}\hat{H}_{\mathrm{eff}}^\dag) + \sum_r \hat{L}_r\hat{\rho}\hat{L}_r^\dag. 
\end{align}
Here the Hermitian Hamiltonian $\hat{H}_0$ and the Lindblad operator $\hat{L}_r$ are given by
\begin{align}
    &\hat{H}_0
    = \sum_{j=1}^{L-1}\bka{\cosh\alpha (\hat{c}_j^\dag\hat{c}_{j+1} + \hat{c}_{j+1}^\dag\hat{c}_j) + U\hat{n}_j\hat{n}_{j+1}}, \nonumber\\
    &\hat{L}_j
    = \sqrt{2\sinh\alpha}(\hat{c}_j+i\hat{c}_{j+1})
    \quad
    (j=1,2,\ldots,L-1), \nonumber\\
    &\hat{L}_0
    = \sqrt{2\sinh\alpha}\hat{c}_1,
    \quad
    \hat{L}_L
    = \sqrt{2\sinh\alpha}\hat{c}_L,
    \label{Seq:H&L}
\end{align}
which lead to the effective Hamiltonian 
\begin{align}
    \hat{H}_{\mathrm{eff}}
    = \hat{H}_0 - \frac{i}{2}\sum_{r}\hat{L}_r^\dag\hat{L}_r 
    = \sum_{j=1}^{L-1} \bka{e^\alpha\hat{c}_j^\dag\hat{c}_{j+1} + e^{-\alpha}\hat{c}_{j+1}^\dag\hat{c}_j + U\hat{n}_j\hat{n}_{j+1}}
    - 2i\sinh\alpha\hat{N}.
\end{align}
We first discuss how to realize the above system.
We consider ultracold fermionic gases on an optical lattice.
Then, we can obtain
the nearest neighbor density-density interaction in $\hat{H}_0$ via 
the dipole-dipole couplings reported in fermionic KRb molecules \cite{Ni2008,Chotia2012,Moses2015}, fermionic and bosonic Dy atoms \cite{Lu2012}, and fermionic NaK molecules \cite{Park2015}.
Following a theoretical proposal in Appendix F of Ref.~\cite{Gong2018}, we also get 
the one-body losses $\hat{L}_j\propto \hat{c}_j+i\hat{c}_{j+1}$ on the basis of reservoir engineering.
Combining these techniques, we prepare the experimental platform for the open quantum system in Eq.~\eqref{Seq:H&L}.
%Note that, although we use the spinless fermion in Eq.~\eqref{Seq:H&L}, we can substitute the spinful or flavored fermion for it under the condition that the coupling among spin or flavor sectors is negligible.

Next, we propose a protocol for measuring the norm dynamics $\|e^{-i\hat{H}_{\rm eff}t}\|$, which exhibits the transient slow dynamics as mentioned in the main text.
The goal in the protocol is to obtain the survival probability $P(t,\hat{\rho}_{\rm init})$.
The intended protocol is as follows.
\begin{enumerate}
    \item Prepare an initial pure state $\ket{\Psi_{\rm init}}$ with the total particle number $N$.
    \item Introduce the density-density interaction and the dissipation in Eq.~\eqref{Seq:H&L} and develop the quantum state under the Lindblad equation in Eq.\eqref{Seq:Lindblad}. 
  %  As mentioned above, the interaction and the dissipation could be realized in ultracold molecules on the basis of dipole-dipole couplings and reservoir engineering, respectively.
    \item After the time interval $t$, turn off the dissipation and count the total particle number of the system.
 %   Note that the continuous measurement or the post-selection is unnecessary in this process.
    In many-body ultracold gases, one can employ the quantum gas microscope technique \cite{Bakr2009} to count the total number of the surviving particles on the optical lattice.
    \item Repeat the procedure above (step 2 to 3) for a fixed $\ket{\Psi_{\rm init}}$ and obtain the survival probability $P(t;\hat{\rho}_{\rm init}=\ket{\Psi_{\rm init}}\bra{\Psi_{\rm init}})$ as the ratio of the trials where the total particle number does not change after the interval $t$ to all the trials.
    \item Change the time interval $t$ into various values, and then evaluate $P(t;\ket{\Psi_{\rm init}}\bra{\Psi_{\rm init}})$ as a function of $t$.
\end{enumerate}
%, one gets the dynamics of the survival probability $P(t;\ket{\Psi_{\rm init}}\bra{\Psi_{\rm init}})$ experimentally.

We discuss the optimal initial pure state $\ket{\Psi_{\rm init}}$ to observe the slow dynamics by the Fock space skin effect.
From Eq.(\ref{eq:surviving_prob}) in Sec. \ref{sec:prob}, the survival probability obeys
\begin{align}
P(t;|\Psi_{\rm init}\rangle\langle \Psi_{\rm init}|
)&=
\tr\left(e^{-i\hat{H}_{\rm eff}t}
|\Psi_{\rm init}\rangle
\langle \Psi_{\rm init}|
e^{i\hat{H}_{\rm eff}t}
\right)    
\nonumber\\
&=\|e^{-i\hat{H}_{\rm eff}t}
|\Psi_{\rm init}\rangle\|^2
\nonumber\\
&\le \|e^{-i\hat{H}_{\rm eff}t}\|^2.
\label{eq:probmax}
\end{align}
The equality in the last equation holds if we choose the right singular vector associated with the maximum singular value of $e^{-i\hat{H}_{\rm eff}t}$ as an initial state: Let $\ket{\Psi_{\max}(t)}$ and $|\Psi_{\max}(t)\rangle\!\rangle$ be the right- and left-singular vectors associated with the maximum singular value of the operator $e^{-i\hat{H}_{\rm eff}t}$. Then because $\|e^{-i\hat{H}_{\rm eff}t}\|$ is the maximum singular value of $e^{-i\hat{H}_{\rm eff}t}$, 
we have
\begin{gather}
    e^{-i\hat{H}_{\rm eff}t}\ket{\Psi_{\max}(t)}
    = \|e^{-i\hat{H}_{\rm eff}t}\| |\Psi_{\max}(t)\rangle\!\rangle,
%    \\
%    \bka{e^{-i\hat{H}_{\rm eff}t}}^\dag|\Psi_{\max}(t)\rangle\!\rangle
%    = \|e^{-i\hat{H}_{\rm eff}t}\| \ket{\Psi_{\max}(t)},
%    \\
\quad
    \norm{\ket{\Psi_{\max}(t)}}
    = \norm{|\Psi_{\max}(t)\rangle\!\rangle}
    = 1,
\end{gather}
which leads to equality in the last equation in Eq.(\ref{eq:probmax}). 
Thus, we can optimize the survival probability at $t$ by choosing $\Psi_{\rm max}(t)$ as the initial state.

\begin{figure}
    \centering
    \includegraphics[width=0.6\linewidth,clip]{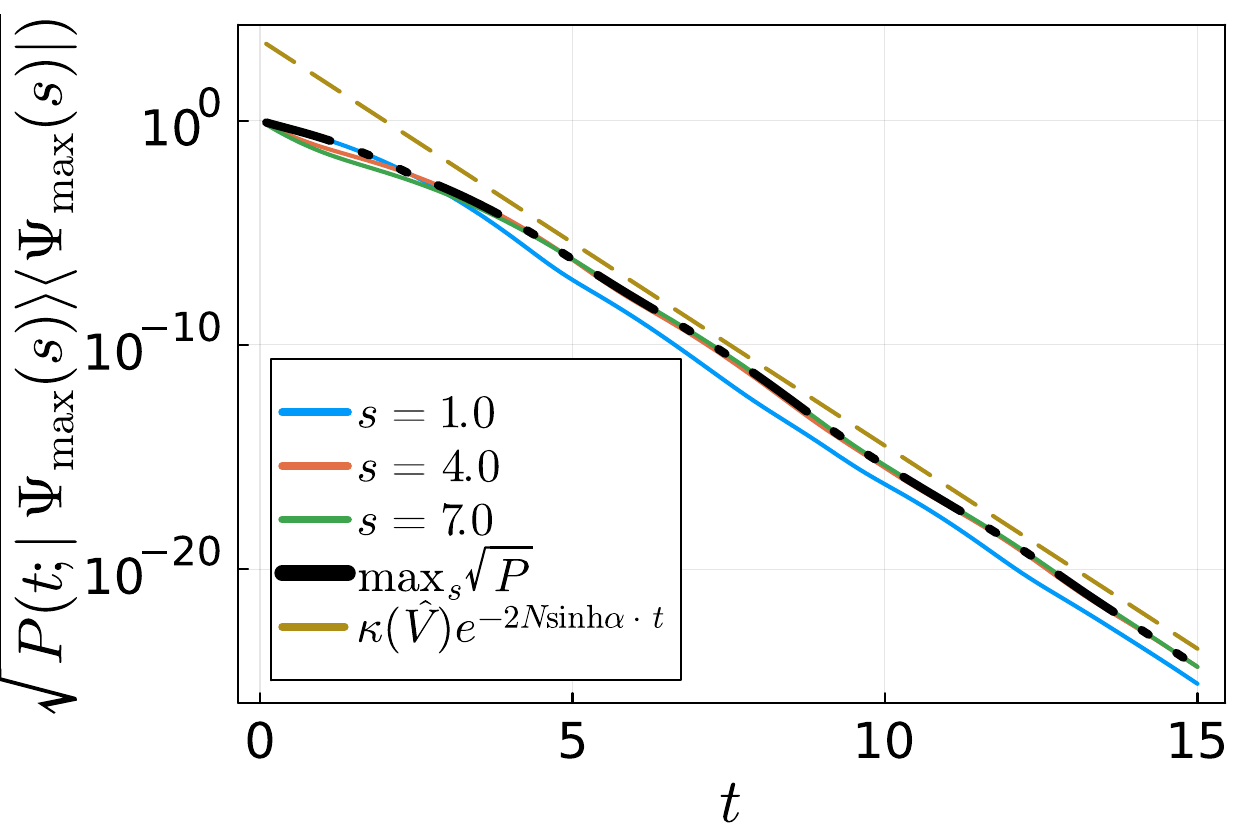}
    \caption{The time evolution of the square root of the survival probabilities $P(t;\ket{\Psi_{\max}(s)}\bra{\Psi_{\max}(s)})$ for $s=1,4,7$ (the blue, orange, and green solid lines, respectively) and the curve connecting the maximum value of them at each $t$ (the black dashed line). We use the model in Eqs.~\eqref{Seq:Lindblad} and \eqref{Seq:H&L} [$\alpha=0.5$, $U=-1.0$, $N=L/2=4$ (half-filling)]. 
    The brown dashed line is the upper bound $\kappa(\hat{V})e^{-2Nt\sinh\alpha}$ of $\|e^{-i\hat{H}_{\rm eff}t}\|$.}
    \label{Sfig:approx_norm_dynamics}
\end{figure}

Now consider the initial state $\Psi_{\rm max}(s)$, which saturates the inequality in Eq. (\ref{eq:probmax}) at $t=s$.
Then, as we have
\begin{align}
\sqrt{P(t;\ket{\Psi_{\rm max}(s)}\bra{\Psi_{\rm max}(s)})}\le \|e^{-i\hat{H}_{\rm eff}t}\|,    
\end{align}
where the equality holds at $t=s$,  $\|e^{-i\hat{H}_{\rm eff}t}\|$
is an envelope of curves for $\sqrt{P(t;\ket{\Psi_{\rm max}(s)}\bra{\Psi_{\rm max}(s)})}$ 
with all possible $s>0$.
The determination of the envelope requires an infinite number of initial states strictly speaking, but in practice, we only need a few initial states. 
Since the slow dynamics are manifest only until the order of the relaxation time $\tau$, as discussed in the main text, the curve connecting the maximum of $\sqrt{P(t;\ket{\Psi_{\rm max}(s)}\bra{\Psi_{\rm max}(s)})}$ for a few different $s\le \mathcal{O}(\tau)$ well approximates the envelope $\|e^{-i\hat{H}_{\rm eff}t}\|$, as illustrated in Fig.~\ref{Sfig:approx_norm_dynamics}. 
The curve shows a deviation from the upper bound $\kappa(\hat{V})e^{-Nt\sinh\alpha}$ until $t\sim 4$, indicating the slow dynamics.
Here, we would like to remark that an exact preparation of the optimal initial state $\Psi_{\rm max}$ is unnecessary in this protocol. If an initial state $\Psi_{\rm init}$ substantially overlaps with $\Psi_{\rm max}$, then one can use it as the optimal initial state.
One can check the overlap with $\Psi_{\rm max}$ by numerical simulations.

Finally, we comment on the experimental relevance of measuring the survival probability.
As shown in Fig.\ref{Sfig:approx_norm_dynamics}, the survival probability eventually decays exponentially.
Still, we can observe the Fock space skin effect-induced slow dynamics of the survival probability for moderate system size.
This is because the Fock space skin effect slows the earlier time dynamics of the survival probability, not a later one.
Thus, the effect can be manifest in experiments before the system fully decays.

\end{document}